\documentclass[a4paper,fleqn,usenatbib]{mnras}
\pdfoutput=1

\usepackage[T1]{fontenc}
\usepackage{ae,aecompl}

\usepackage{graphicx}	
\usepackage{amsmath}	
\usepackage{amssymb}	
\usepackage{adjustbox}
\usepackage{tablefootnote}

\title[Dust Attenuation in Clumpy, Star-Forming Galaxies at 0.07 < $z$
< 0.14]{Integrated and Resolved Dust Attenuation in Clumpy
  Star-Forming Galaxies at 0.07 < $z$ < 0.14}

\author[R. Bassett et al.]{
Robert Bassett,$^{1,2}$\thanks{E-mail: robert.bassett@uwa.edu.au (ICRAR)}
Karl Glazebrook,$^{1}$
David B. Fisher,$^{1}$
Emily Wisnioski,$^{3}$\newauthor
Ivana Damjanov,$^{4}$
Roberto Abraham,$^{5}$
Danail Obreschkow,$^{2}$\newauthor
Andrew W. Green,$^{6}$
Elisabete da Cunha,$^{7}$
and Peter J. McGregor$^{7}$
\\
$^{1}$Centre for Astrophysics and Supercomputing, Swinburne University
of Technology, Hawthorn, Australia\\
$^{2}$International Centre for Radio Astronomy Research, University of
Western Australia, 7 Fairway, Crawley, WA 6009, Australia\\
$^{3}$Max-Planck-Institut f{\"u}r extraterrestrische Physik (MPE),
Scheinerstrasse 1, D-85748 Garching, Germany\\
$^{4}$Harvard-Smithsonian Center for Astrophysics, 60 Garden Street,
Cambridge, MA 02138, USA\\
$^{5}$Department of Astronomy and Astrophysics, University of
Toronto, 50 St George Street, Toronto, ON M5S 3H8, Canada\\
$^{6}$Australian Astronomical Observatory, PO Box 970, North Ryde, NSW 1670, Australia\\
$^{7}$Research School of Astronomy \& Astrophysics, The Australian
National University, Cotter Rd, Weston, ACT 2611, Australia
}

\date{Accepted XXX. Received YYY; in original form ZZZ}

\pubyear{2016}

\begin{document}
\label{firstpage}
\pagerange{\pageref{firstpage}--\pageref{lastpage}}
\maketitle

\begin{abstract}
Dust attenuation in galaxies has been extensively studied nearby, however,
there are still many unknowns regarding attenuation in distant galaxies. We 
contribute to this effort using observations of star-forming galaxies in the redshift range
z = 0.05-0.15 from the DYNAMO survey
\citep{green14}. Highly star-forming DYNAMO galaxies share many
similar attributes to clumpy, star-forming galaxies at high redshift. Considering integrated
Sloan Digital Sky Survey \citep{york00} observations, trends between attenuation
and other galaxy properties for DYNAMO galaxies are well matched to
star-forming galaxies at high redshift. Integrated gas attenuations of DYNAMO galaxies are 0.2-2.0 mags in the V-band, and the ratio of $E(B-V)_{stars}$ and $E(B-V)_{gas}$
is 0.78-0.08 \citep[compared to 0.44 at low redshift,][]{calz97}. Four highly star-forming
DYNAMO galaxies were observed at H$\alpha$ using the Hubble Space Telescope and at
Pa$\alpha$ using integral field spectroscopy at Keck. The latter achieve similar resolution
($\sim$0.8-1 kpc) to our HST imaging using adaptive optics, providing resolved observations of
gas attenuations of these galaxies on sub-kpc scales. We find < 1.0 mag of variation
in attenuation (at H$\alpha$) from clump to clump, with no evidence of highly attenuated star
formation. Attenuations are in the range 0.3-2.2 mags in the V band, consistent with
attenuations of low redshift star-forming galaxies. The small spatial variation on attenuation suggests that a majority of the
star-formation activity in these four galaxies occurs in relatively unobscured regions and, thus,
star-formation is well characterised by our H$\alpha$ observations.
\end{abstract}




\section{Introduction}\label{section:introduction}

Understanding dust attenuation in galaxies is an essential ingredient
in studies of galaxy formation and evolution. In the current paradigm,
the relationship between star-formation rate (SFR) and stellar
mass \citep[the star-forming ``main sequence'', e.g.][]{brinchmann04,elbaz07,whitaker12} is
an important tool, however, it is highly dependent on prescriptions
for dust attenuation due to the strong influence of dust on various SFR
indicators. An accurate description of the effects of
dust during the peak of cosmic star-formation \citep[$1 < z <
3$,][]{lilly96,madau98,reddy09,cucciati12} remains elusive due to observational 
constraints. Star-forming galaxies at these epochs are found to form
stars much more rapidly than local galaxies, and it is unclear whether
or not local attenuation relations \citep[see][for a review]{calz01}
can be applied \citep{price14,reddy15}. The assumption of a
screen-like dust geometry, as well as the
relationship between attenuation and mass \citep{garn10b}, may 
affect the normalisation of the star-forming main-sequence, both of
which are difficult to study at high redshift. 

The flux of the H$\alpha$
emission line in the optical is a standard indicator of
the SFR of a given galaxy \citep{kennicutt12} and, although it can be
significantly attenuated by dust, it is commonly used
to study star-formation at both low and high redshift. For an ideal correction, one
would measure the attenuation for the gaseous component by comparing
the H$\alpha$ to H$\beta$ recombination line ratio \citep[$R($H$\alpha$,H$\beta)$, the Balmer
decrement,][]{berman36,mathis83} to the intrinsic value for typical
star-forming regions \citep[$R($H$\alpha$,H$\beta)_{int} =2.87$, from
Case B recombination][]{hummer87,osterbrock89}. By definition, this
indicator is directly associated with line emission originating from
star-forming regions. Detecting both
H$\alpha$ and H$\beta$ at $z>1$ for single galaxies, however, is
difficult due to the faintness of H$\beta$
\citep[particularly at high
attenuation,][]{erb06c,yoshikawa10,dominguez13,price14}. Furthermore
using current instruments this may require observations in multiple
bands \citep[e.g.][]{kashino13}. For these reasons, the
attenuation of the stellar light may be measured first, often through fitting of
the spectral energy distribution (SED), with the attenuation of
H$\alpha$ then estimated based on the local relation for starburst
galaxies: $E(B-V)_{stars} = 0.44 \times E(B-V)_{gas}$
\citep{calz97,yoshikawa10,mancini11}. 
The universality of
this relation is far from certain, particularly at high redshift
\citep{wild11,kashino13,wuyts13,price14,reddy15,battisti16,puglisi16}. Furthermore,
\citet{kreckel13} show that the applicability of the local relation
can vary significantly within individual galaxies. Thus further investigation
of stellar versus ionized gas attenuation is warranted.

The work of \citet{kreckel13} shows that studying attenuation in
external galaxies is further complicated by dust geometry, which is
often only broadly characterised due to the integrated nature of
many observations. Within the Milky Way, variations in extinction
from 0 to 30 magnitudes have been observed \citep{cambr11} while
integrated attenuations of face-on disk galaxies at $z<0.8$ typically
fall in the narrow range of $\sim0.5-2.0$ magnitudes in the B-band
\citep[e.g.][]{keel01,matthews01,takeuchi05b,cortese08}. The
difference in measurements of the Milky Way versus external galaxies
highlights the limitations of integrated attenuation measurements. A solution to
this is offered by resolved observations such as those of 
\citet{calzetti97} who target NGC 5253 finding variation in attenuation from 0 to 9
mag. With the recent proliferation of integral field spectroscopy
(IFS) resolved observations of attenuation such as these are becoming more commonplace
\citep[e.g.][]{bedregal09,kreckel13,piqueras13}. 

The effects of large-scale dust geometry will likely play an important role in highly star-forming
galaxies at high redshift hosting massive star-forming clumps
\citep{wright09,forster09,wisnioski11,epinat12,swinbank12}. Although the exact
definition of a star-forming clump varies somewhat between authors, in
general this terminology refers to individual, large-scale,
star-forming regions typically identified by their strong emission lines.
Currently little is known regarding clump-to-clump variations in
attenuation although one study by \citet{genzel13} infers possible
attenuations of up to 50 mags, albeit with uncertain assumptions
regarding dust geometry, based on molecular gas observations. Locally,
the most highly attenuated galaxies are highly
star-forming, infrared (IR) bright galaxies known as luminous (and
ultraluminous) star-forming galaxies \citep[LIRGS/ULIRGS see][for a
review]{sanders96}. Integrated attenuations of these objects are
measured to be $\sim$3-5.5 mag in the $V$-band \citep[e.g.][]{calz05,alonso06}. These
values fail to give the full picture, however, as they assume simple
dust geometries while resolved observations reveal pixel to pixel
variation of $\sim$1-20 mag \citep[e.g.][]{piqueras13}. The
counterpart at high redshift are so called submillimeter galaxies
\citep[SMGs; see][for a review]{blain02} that radiate a very large
fraction of their energy in the IR, possibly implying that they
contain enough dust to be nearly optically thick. Current results
suggest galaxies on the high redshift star forming main sequence are
significantly less attenuated, though, with values closer to local
main sequence galaxies \citep[e.g.][]{price14,reddy15}. However,
considering the results of \citet{alonso06} and \citet{piqueras13}
for local ULIRGS, further study of dust geometry in clumpy,
star-forming galaxies is necessary.

In this paper we highlight two
important open questions regarding dust geometry in
clumpy, high redshift galaxies. Are clumps a genuine morphological feature, or are variations
in dust attenuation a large contributor to the appearance of these
galaxies? And, if dust is highly variable spatially, will this strongly
bias measurements of galaxy SFR at high redshift?
Due to the difficulties in studying attenuation at high
redshift highlighted above, studies of dust geometry
using resolved IFS observations of extremely star-forming galaxies at
low redshift may present a viable step forward.

Here we explore the spatially resolved
attenuation properties of a unique sample of star-forming galaxies
taken from the DYnamics of Newly Assembled Massive Objects survey
\citep[DYNAMO,][]{green14}. DYNAMO galaxies are drawn from the Sloan
Digital Sky Survey Data Release 4 \citep[SDSS DR4,][]{york00} and
previous observations have shown highly star-forming DYNAMO galaxies
to share many similar properties to clumpy galaxies at high
redshift
\citep{green14,bassett14,fisher14,obreschkow15}. 
Here we compare Hubble Space Telescope (HST) H$\alpha$ photometry and adaptive optics
(AO) assisted infrared (IR) IFS of Paschen $\alpha$
(Pa$\alpha$) to study the spatially resolved gas attenuation
in four clumpy, high SFR DYNAMO galaxies on $\sim1$ kpc
scales.
Although kpc scale observations of
clumpy galaxies at $z>1.5$ are possible using
gravitational lensing and/or AO, obtaining such
observations of both H$\alpha$ and H$\beta$
at these redshifts has to this point been infeasible. Thus, our study provides a necessary
test of clump-to-clump variation in attenuation for turbulent, clumpy
galaxies.

This paper is laid out as follows: in Section \ref{section:ch4data} we
present the data sets used for both our integrated and resolved
analyses, in Section \ref{section:intav} we present a brief analysis of
the integrated stellar and gas attenuation properties of DYNAMO galaxies, in Section \ref{section:ch4resavmeas} we describe the
analysis performed in producing resolved maps of attenuation for our
subsample of four extreme DYNAMO galaxies, in Section
\ref{section:ch4resavres} we present the results of our resolved
attenuation analysis, in Section \ref{section:ch4discussion} we
provide a discussion of all of our results, and in Section
\ref{section:ch4summary} we give a brief summary and itemise our
conclusions. Throughout this work, we adopt a flat cosmology with
$H_{0}=70$ km s$^{-1}$ Mpc$^{-1}$ and $\Omega_{M}=0.3$.

\section{Samples and Data}\label{section:ch4data}

\subsection{DYNAMO: A Local Sample of High-$z$ Analogs}

\begin{figure}
  \includegraphics[width=3.5in]{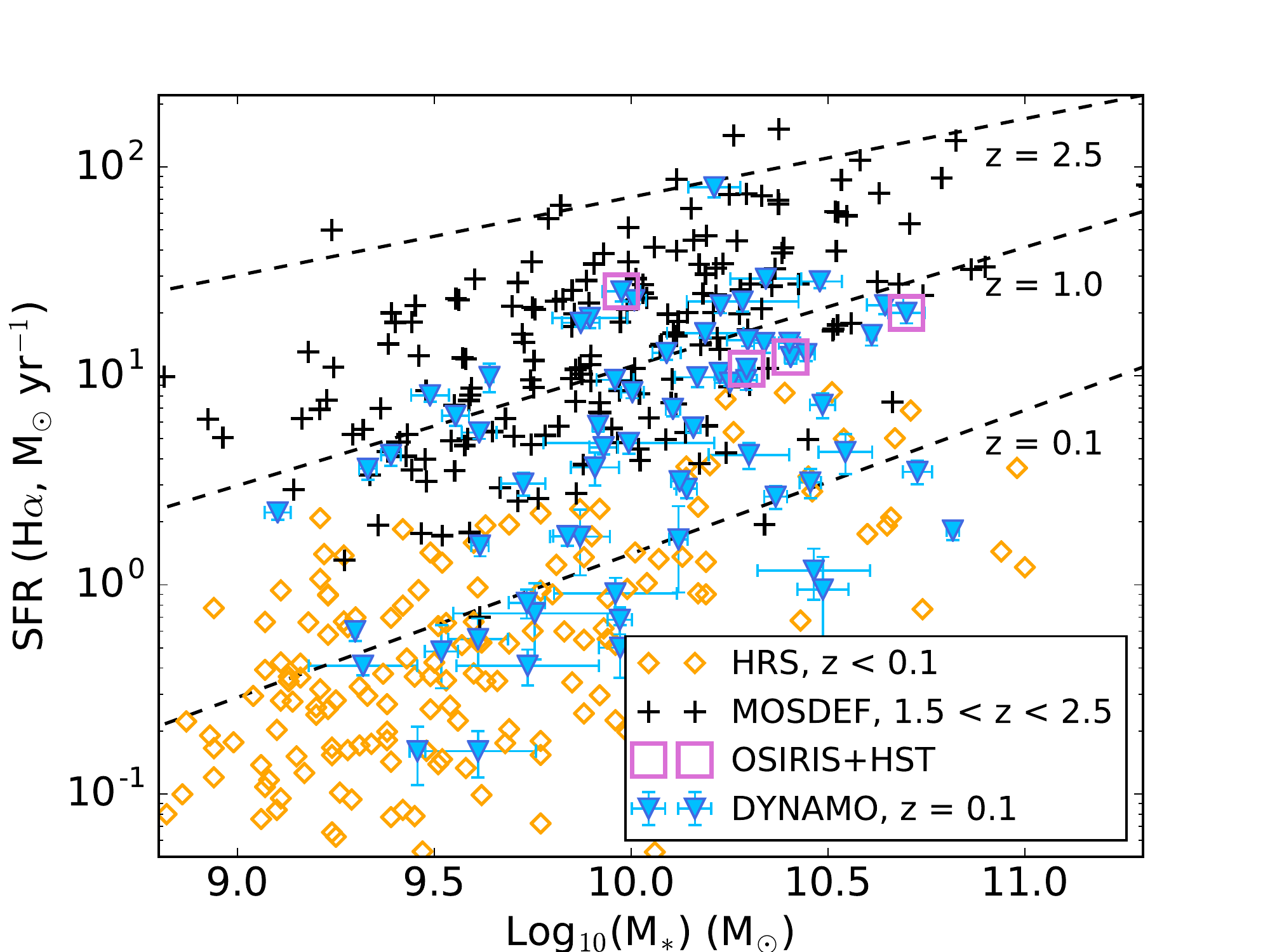}
  \vspace{-2em}
  \caption{$M_{*}$ vs SFR (from H$\alpha$ luminosity) for 67 DYNAMO
    galaxies from \citet{green14} along with our two
    comparison samples. At low redshift we select late-type star-forming galaxies from the Herschel Reference Survey while the
    MOSDEF survey provides a sample of star-forming galaxies at
    $z>1.5$. We also plot the redshift evolution of the main-sequence
    of star-forming galaxies given by \citet{whitaker12} at $z$=0.1,
    1.0, and 2.5. DYNAMO galaxies exhibit significant overlap between
    both samples, bridging the gap in star-formation properties from
    low to high redshifts. The four galaxies for which resolved
    observations are also presented in this paper
    are indicated by open purple squares.}\label{figure:ms_plot}
\end{figure}

DYNAMO galaxies were selected in two redshift bins
chosen to avoid contamination of the H$\alpha$ emission line by common
night sky emission lines. In both bins active galactic nuclei (AGN)
have been excluded using the standard procedure of \citet*[][the
``BPT diagram'']{baldwin81} in order to focus on purely star-forming,
line-emitting galaxies.
Initial IFS observations revealed a number of galaxies
with clumpy disk morphologies
and turbulent kinematics \citep[as indicated by a large gas velocity dispersion,
$\sigma_{gas}$,][]{green14} reminiscent of star-forming galaxies observed at
$z>1$. Follow-up observations using IFS at the Gemini and Keck observatories have confirmed
that this kinematic signature is not an artifact of resolution for a
handful of
DYNAMO disks at $z=0.1$ \citep[][Oliva-Altamirano et al. \textit{in prep}]{bassett14}. Three DYNAMO galaxies
have been securely detected in unresolved CO observations indicating
gas fractions of up to 30\% \citep[measured for DYNAMO G 04-1 included
in this work,][]{fisher14}. For these reasons a
subset of highly star-forming DYNAMO galaxies represent the best-known
sample at $z<0.2$ for studying star-formation processes more common
at $z>1.5$.

The relatively wide range of properties for the DYNAMO sample is
reflected in Figure \ref{figure:ms_plot} where we show the relationship between
$M_{*}$ and SFR for DYNAMO
galaxies. $M_{*}$ values are taken from the MPA-JHU VAC, and they are
calculated using the population synthesis methods presented in
\citet{kauffmann03a}. SFR is calculated based on the spatially
integrated H$\alpha$ luminosity from IFS observations presented in
\citet{green14} with an attenuation correction based on the Balmer
decrement from SDSS observations \citep{tremonti04} and following the
method of \citet{calz97}. The attenuation corrected H$\alpha$
luminosity is then convrted to SFR following calibration of \citet{kenni98} modified
for a \citet{chab03} initial mass function. Here we compare DYNAMO galaxies with the $M_{*}$-SFR
relationships for low redshift and high redshift star-forming
galaxies. These redshift regimes are represented by the Herschel Reference Survey (HRS) of nearby
star-forming galaxies \citep{boselli10,cortese12,boselli15} and the MOSFIRE Deep Evolution Field
(MOSDEF) survey \citep[][see Section \ref{section:compsamp} for further
description of the MOSDEF sample]{kriek15,reddy15} respectively. Overplotted as dashed
lines are fits of the main sequence of star-forming galaxies
\citep[e.g][]{brinchmann04} taken from \citet{whitaker12} at redshifts
0.1, 1.0 and 2.5. DYNAMO galaxies are
found to exhibit a large scatter in SFR at fixed $M_{*}$ with galaxies
typical of the local and $z>1.0$ main sequence represented. We
indicate galaxies for which resolved attenuations are explored in this
work with open purple squares. These four galaxies are found to have
SFR and $M_{*}$ roughly consistent with the $z=1.0$ main sequence.

\subsection{Unresolved Sample}\label{section:intavdata}

We begin this work by briefly investigating the integrated
attenuations of the full sample of DYNAMO galaxies presented by
\citet{green14} in order to provide context for our 
resolved observations. These 67 galaxies make up our unresolved
attenuations sample. We show the relationship between $M_{*}$ and SFR
for these 67 galaxies as inverted blue triangles in Figure
\ref{figure:ms_plot}. This shows that DYNAMO galaxies exhibit a wide
range in star-formation properties.

Of particular interest here is the comparison between attenuations
suffered by the stellar and ionized gas components of DYNAMO
galaxies. If highly star-forming DYNAMO galaxies are true analogs of
high redshift galaxies, they may exhibit a similar departure from local
galaxies that show roughly two times more attenuation for ionised
gas \citep[e.g.]{calz97,price14,reddy15}. Our method of calculating
integrated attenuations based on the Max Planck Institute for
Astrophysics and Johns Hopkins University Value Added Catalog
\citep[MPA-JHU VAC;][]{kauffmann03a,tremonti04} is described in Section
\ref{section:intdgs}. 

\subsection{Four Resolved Attenuations Galaxies}

\begin{figure*}
  \includegraphics[width=\textwidth]{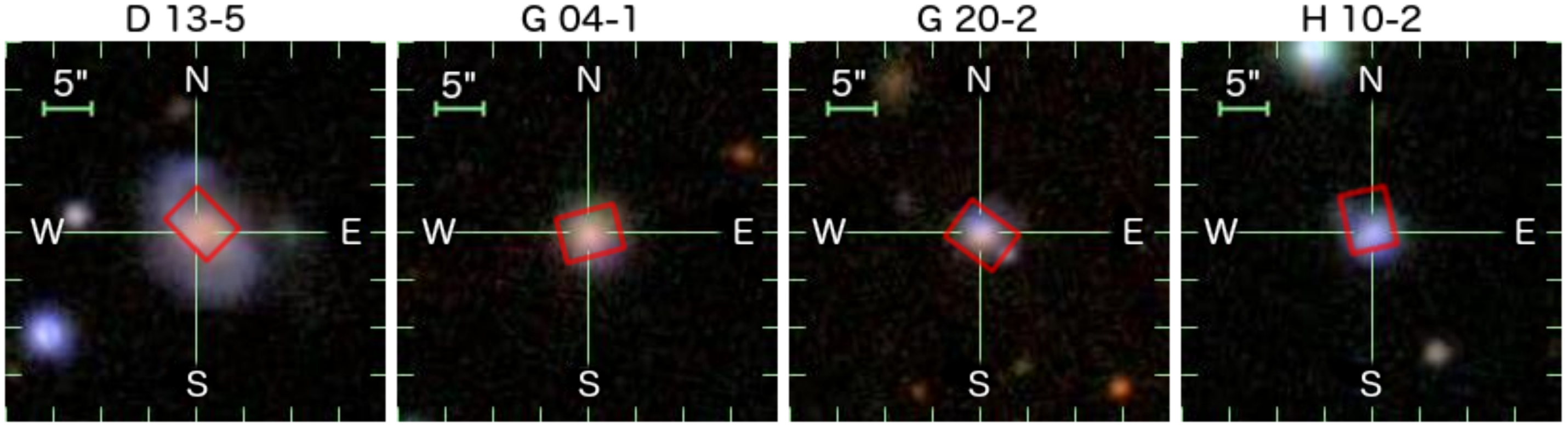}
  \caption{SDSS postage stamps of the four galaxies in our resolved
    attenuation sample. In each panel a scale bar indicating a size of
  5$\farcs$0 is shown in green and the approximate field of view of
  our OSIRIS observations is shown as a red rectangle. Note that the
  field of view of our HST observations is significantly larger than
  the postage stamps pictured here.}\label{figure:sdssstamps}
\end{figure*}

Galaxies in our resolved attenuations sample are selected from the
overlap of two separate DYNAMO programs, a HST photometric program
targeting H$\alpha$ in 10 galaxies (see Section \ref{section:ch4HSTd})
and an AO assisted IR IFS program using the OH-Suppressing Infrared Integral field
Spectrograph \citep[OSIRIS,][]{larkin06} at the Keck observatory
targeting Pa$\alpha$ in 15 galaxies (see Section
\ref{section:ch4OSIRISd}). These programs were designed to explore
the properties of analogs to high redshift, clumpy galaxies, thus both
selections include primarily highly star-forming DYNAMO galaxies. The primary
goals of our H$\alpha$ and Pa$\alpha$ programs were to study the
size-luminosity relation for clumps in DYNAMO galaxies \citep{fisher16} and sub-kpc kinematics of DYNAMO galaxies
(Oliva-Altamirano et al. in prep.) respectively. 

Galaxies
observed in each program were subject to different constraints,
therefore only four galaxies, D 13-5, G 04-1, G 20-2, and H 10-2, comprise the overlap between
our HST and OSIRIS programs.
These four DYNAMO galaxies  were previously classified as
rotating disks by \citet{green14}, and they exhibit large velocity H$\alpha$
velocity dispersion (45-60 km s$^{-1}$), and form stars rapidly
(SFR$\sim$15-40 M$_{\odot}$ yr$^{-1}$). Integrated galaxy properties
for this subsample are summarised in Table
\ref{table:ch4physparam} and SDSS postage stamps are shown in Figure
\ref{figure:sdssstamps} with red rectangles indicating the approximate
field of view of our OSIRIS observations.

\begin{table}[t]
  \centering
  \begin{adjustbox}{max width=\columnwidth}
  \vspace{2mm}
  \begin{tabular}{ c c c c c c c }
    \hline\hline
    ID & \textit{z} & $\mathcal{M}_{*}$\tablefootnote{Stellar mass from SED
    fitting \citep{kauffmann03a} scaled by 0.88 to convert to
    \citet{kroupa01} initial mass function.} &
  SFR$_{H\alpha}$\tablefootnote{SFR measured from H$\alpha$ IFS
    observations of \citet{green14} corrected for attenuation based on
                                               SDSS Balmer decrement
                                               following the method of
                                               \citet{calz97}} &
  $\sigma_{gas}$\tablefootnote{H$\alpha$ velocity dispersion from disk fit
    models of \citet{green14}} & PSF Scale\tablefootnote{Based on the core
    size of our Keck-AO PSF measured through 2D Gaussian fitting to
    observed standard stars. The value in parenthesis reflects the
                                 spatial scale based on the optical
                                 seeing, which is four times larger.}
    & Type\\
    &  & ($10^{9}$ M$_{\odot}$) & (M$_{\odot}$yr$^{-1}$) & (km
                                                           s$^{-1}$) &
                                                                       (kpc) &\\
    \hline
    D 13-5 & 0.075 & 53.84 & 12.31$\pm$0.86 & 46 & 0.21 (0.85) & disk\\
    G 04-1 & 0.129 & 64.74 & 20.00$\pm$2.17 & 50 & 0.35 (1.39) & disk\\
    G 20-2 & 0.141 & 21.56 & 10.80$\pm$0.66 & 45 & 0.38 (1.50) & disk\\
    H 10-2 & 0.149 & 9.5 & 25.35$\pm$2.68 & 59 & 0.39 (1.57) & merger\\ 
    \hline
  \end{tabular}
  \end{adjustbox}
    \caption{Resolved Attenuation Sample: Previous DYNAMO
    Observations}
    \label{table:ch4physparam}
\end{table}

Galaxies D 13-5, G 04-1, and G 20-2 are characterised by undisturbed
continuum morphologies and smooth, disk-like rotation. Two of these,
G 04-1 and G 20-2, are in the $z\sim0.1$ and high H$\alpha$ flux bins of
the original DYNAMO sample and as such their angular sizes and fluxes
are well suited to observation with OSIRIS. We also note that the
stellar versus ionized gas kinematics in galaxies
G 04-1 and G 20-2 were the subject of another recent work
\citep{bassett14} finding these galaxies to be consistent with
rotating disks using higher resolution IFS observations from the
Gemini Observatory. The third disk-like
galaxy, D 13-5, is at a lower redshift and we therefore primarily cover
only the central most regions of this galaxy, missing flux at large
radii.

The fourth galaxy in our sample, H 10-2, was originally
identified as an extremely H$\alpha$ luminous galaxy with disk-like
rotation from our initial observations. Deep optical IFS using Gemini
MultiObject Spectrometer (GMOS)
at The Gemini Observatory \citep[see][for description of GMOS observations]{bassett14} revealed a
second component rotating at 90 degrees relative to the previously
identified kinematic axis. These kinematic components correspond to two significant peaks in
continuum emission leading us to reclassify this galaxy as an ongoing
merger. Regardless, we include this object in our analysis, as it is
valuable as a comparison to our disk sample.

\subsubsection{H$\alpha$ Photometric Observations}\label{section:ch4HSTd}

Our H$\alpha$ photometry was collected using HST Advanced Camera for
Surveys Wide-Field Camera
(ACS/WFC) FR647M
(Proposal ID 12977, PI: Damjanov). Observations were performed
using the FR716N and FR782N ramp filters, which are equivalent to
tunable narrow-width pass band filters, targeting H$\alpha$
emission with a 2\% bandwidth. Continuum
subtraction is achieved using observations in the associated continuum
filter, FR647M. Integration times for our H$\alpha$ and continuum
observations are 45 min and 15 min respectively. The typical HST
pipeline is used to reduce images for analysis. 
The full HST sample of 10 detected galaxies is
presented in \citet{fisher16}.

\subsubsection{Infrared IFS Observations and Reduction}\label{section:ch4OSIRISd}

Pa$\alpha$ data comes from IFS observations using the OSIRIS
instrument at the Keck Observatory. OSIRIS is a lenslet array spectrograph
with a $2048 \times 2048$ Hawaii-2 detector and spectral resolution R $\sim$ 3000
in the 100 mas spatial scale. We perform our observations with
the aid of natural guide star (NGS) or laser guide star (LGS) AO
systems at Keck \citep{wizinowich06,vandam06}. Galaxy D 
13-5 was observed on Keck I in July 2012 using NGS-AO ($\sim$0$\farcs$85
optical seeing), G 20-2 was also observed in July 2012 on Keck I
with the aid of LGS-AO ($\sim$0$\farcs$65 optical seeing), G 04-1 was
observed on Keck I in September 2012 using LGS-AO ($\sim$0$\farcs$60
optical seeing), and finally galaxy H
10-2 was observed at Keck II in March 2010 using LGS-AO. Note that
while the AO point spread function (PSF) has a core that is
significantly narrower than the seeing (typically $\sim$0$\farcs$15 in
our observations), the shape and width are known to vary
significantly both from night to night as well as within a single
night of observations. Discussion of the AO PSF is revisited in Section
\ref{section:ch4psfmatch}. We will present the analysis of
high-resolution Pa alpha kinematic properties for star-forming clumps
in the full sample of 15 DYNAMO galaxies observed with OSIRIS in the
upcoming publication (Oliva-Altamirano et al. in prep) 

The standard observing procedure was as follows. We first acquired the
tip-tilt star and applied the optimal position angle of OSIRIS to
position the star within the unvignetted field-of-view of the LGS-AO
system. Short, 60s, integrations were taken on the star for
calculations of the PSF and to centre the star in the field for the
target offsets. Multiple
positions for science observations were dithered by 0$\farcs$05 around
the base positions in each exposure to remove bad pixels and cosmic
ray contamination. Sky frames were taken completely offset from the
object as the galaxies fill the whole
field of view of OSIRIS. All galaxies were observed in the 100mas
scale.

Data reduction was completed using the OSIRIS data reduction pipeline
version 2.3, and custom IDL routines developed for faint emission-line
spectra. The pipeline removes crosstalk, detector glitches, and cosmic
rays before it mosaics individual exposures and assembles a reduced
data cube with two spatial dimensions and one spectral
dimension. First order sky subtraction was achieved by the spatial
nodding on the sky. Further sky subtraction was applied using custom
IDL routines that employ the methods of \citet{davies07}. We initially
perform a spectrophotometric flux calibration to our
OSIRIS IFS observations by comparing standard star observations to a
synthetic spectrum of Vega, however, strong variability in the AO PSF
over a night results in large systematic uncertainties on the order of
$\sim$30\% (when comparing observed OSIRIS continuum flux densities to
catalog values from 2MASS), comparable to other works using OSIRIS \citep[e.g.][]{law09}. For this
reason, we subsequently adjust the integrated flux of Pa$\alpha$ for each
galaxy to match the expected flux based on the SDSS Balmer
decrement. This is described further in Section
\ref{section:ch4bdcorr} and the correction varies by a factor of
1.07 for H 10-2 to 2.61 for D 13-5. 

\subsubsection{Ancillary IFS Data}\label{section:ch4ancifs}

\noindent\textbf{G 20-2 [OIII]/H$\beta$:} We also explore variations
in the ionization state of the gas in galaxy G 20-2 using optical GMOS-IFS observations of H$\beta$ and [OIII]
(5007 {\AA}). A full description of the observations and data
reduction can be found in \citet{bassett14}.
Using these observations we can recover some information
on the ionization state of the gas by comparing the strength of the Balmer emission to that
of the forbidden transition of oxygen. In particular, we explore
the possibility of a low-luminosity AGN that may explain irregularity
in the measured ionized gas attenuation in the central regions (see
Section \ref{section:ch4resavres}).

Flux calibration is achieved by matching a spectrum summed in an
approximation of the SDSS fibre footprint on our IFS datacube to the
observed SDSS spectrum in the same wavelength range. We then map both the [OIII] and H$\beta$ emission lines as described in
Section \ref{section:ch4ifselm}. We note that because they are separated by <
150 {\AA} mapping the ratio of these two lines does not depend significantly
on the absolute accuracy of this calibration. A map of
log$_{10}($[OIII]$/$H$\beta)$ is reproduced for G 20-2 in comparison to
$A_{H\alpha}$ (the attenuation at H$\alpha$) in Section
\ref{section:ch4resavres}. GMOS [OIII]/H$\beta$ maps are registered to
match our OSIRIS observations as described in Appendix
\ref{section:ch4imreg}.

The remaining three
galaxies in our resolved sample have also been observed in H$\beta$
and [OIII], however, we find no strong correspondence between the maps
of [OIII]$/$H$\beta$ and attenuation for D 13-5, G 04-1, or H 10-2, perhaps owing to the
lower resolution of these observations (typical optical seeing of
0$\farcs$8-1$\farcs$2). 

\vspace{1mm}

\noindent\textbf{[NII]/H$\alpha$:}
In addition to optical IFS from GMOS, we also have data from the
original DYNAMO observations performed using the AAOmega-SPIRAL and
Wide Field Spectrograph (WiFeS) IFS, which targeted H$\alpha$
emission. The details of these observations and data reduction can be
found in \citep{green14}. SPIRAL and WiFeS observations were taken
with a smaller 
aperture relative to GMOS, with poorer seeing (1\farcs0-1\farcs5), and
with much shorter exposure times. As such, there is little resolved
substructure in these observations, however, they are suitable for
observing any strong radial trends in emission line
properties as well as measuring global properties. 
From the SPIRAL and WiFeS datacubes, we characterise the
relative fluxes of $[NII]$ and H$\alpha$ in each of our galaxies. The
wavelength range of our HST ramp filter is known to contain $[NII]$ in
addition to H$\alpha$, thus we use SPIRAL and WiFeS data to correct
for this. A description of this correction is described in Section
\ref{section:ch4NII_Ha}.

\subsection{High Redshift Comparison: MOSDEF}\label{section:compsamp}

The MOSFIRE Deep Evolution Field \citep[MOSDEF,][]{kriek15} survey is
an ongoing IR spectroscopic
survey of galaxies at $z>1.5$ that is being performed using the
MOSFIRE spectrograph on the Keck I telescope. The goal of this survey is
to explore the evolution in the rest-frame optical spectra of
$\sim$1500 galaxies in the redshift range 1.4 < $z$ < 3.4. 

We compare DYNAMO galaxies with the sample presented
in \citet{reddy15} who focus on dust attenuation for 224 star-forming MOSDEF galaxies
with secure detections of both H$\alpha$ and H$\beta$ at 1.4 < $z$ < 2.6. Galaxies
in this sample are selected to have a roughly constant mass limit of
$\sim$10$^9$ M$_{\odot}$ independent of redshift. Stellar masses and attenuations were
computed through SED fitting of 3D-HST photometry \citep{skelton14}, which covers the
rest-frame UV to near-IR of MOSDEF galaxies. Emission line fluxes are
then measured from MOSFIRE spectra with the best fitting SED model
subtracted similarly to the MPA-JHU measurements, thus accounting for the effect of Balmer absorption on the
observed H$\alpha$ and H$\beta$ fluxes. Ionized gas attenuations are
then computed based on the Balmer decrement. Finally, values of attenuation corrected
H$\alpha$ fluxes are converted to SFR using the calibration of
\citet{kenni98} and assuming a \citet{chab03} IMF. 

We choose to reject
from our analysis those MOSDEF
galaxies lacking secure detections of the H$\beta$ emission line,
leaving a sample of 121 star-forming $z>1.5$ galaxies considered in
this work. MOSDEF represents the current largest sample of $z>1.5$
galaxies with secure H$\alpha$ and H$\beta$ detections making it the
ideal sample for comparing DYNAMO galaxies to typical star-forming
galaxies at high redshift.

\section{Integrated Attenuation Properties of DYNAMO Galaxies}\label{section:intav}

\subsection{Measuring DYNAMO Integrated Attenuations}\label{section:intdgs}

As is customary we work with the observed
quantity $E(B-V)$, the colour excess, which represents the difference
in attenuation between the B and V bands. This quantity is related to
the attenuation through the equation:
\begin{equation}\label{equation:eq1}
R_{V}=\frac{A_{V}}{E(B-V)}
\end{equation}
where $A_{V}$ is the attenuation in the V band, and $R_{V}$ is a
constant that can vary from galaxy to galaxy, but
has a typical value of 4.05$\pm$0.80 for nearby starbursts 
\citep{calz00}. 

Predicting the attenuation at wavelengths beyond the V band requires
the evaluation or assumption of an attenuation curve, $k(\lambda)$.
Various attenuation curves have been measured empirically
and, on average, the shape is exponential from the optical to IR with
more complex behaviour at shorter wavelengths
\citep[][among
others]{stecher65,fitzpatrick86b,card89,calz97,charlot00}. \citet{wild11}
and \citet{reddy15} clearly show that the choice of attenuation curve can have a
direct impact on measurements of galaxy properties such as $M_{*}$ and
SFR. Furthermore, the value of $R_{V}$ in Equation \ref{equation:eq1}
is also known to depend directly on the shape of the chosen
attenuation curve. Exploring the effects of a varying attenuation
curve is beyond the scope of this work, however, thus we simply make
note of this complication here.

Another issue faced by our measurements of integrated attenuations for
DYNAMO galaxies is the small size of the SDSS fibre, which has a
diameter of 3$\farcs$0. This means that our attenuation measurements
will be biased towards the average value observed in the central most
regions of DYNAMO galaxies. A decreasing $A_{H\alpha}$ with increasing
radius has been observed in some galaxies
\citep[e.g.][]{munozmateos09,nelson16} and, if this is the case for
DYNAMO, may result in an overestimate of the average attenuation. We
argue that complex spatial variation in dust attenuation, regardless of whether this
occurs inside or outside of the SDSS fibre footprint,
can not easily be accounted for in integrated measurements,
further highlighting the necessity for spatially resolved
observations of attenuation. Our integrated measurements for ionized gas and
stars are both calculated based on SDSS spectral observations, as
described below, and thus
a comparison between the two is garunteed to be sampling the same
spatial regions of the galaxies. Finally we note that the effect of the
SDSS fibre size will be more pronounced for galaxies in the low
redshift DYNAMO range. We show in Section
\ref{section:ch4resavres} that the sizes of higher redshift DYNAMO galaxies
are well matched to the 3$\farcs$0 diameter.

\subsubsection{DYNAMO $E(B-V)_{gas}$}

We compute $E(B-V)_{gas}$ values based on the ratio of H$\alpha$ to
H$\beta$ fluxes \citep[the ``Balmer decrement'', e.g.][]{berman36,hummer87,osterbrock89} from the MPA-JHU VAC using the
methods of \citet{calz97}. Unlike the sample of
MOSDEF galaxies from
\citet{reddy15}, all DYNAMO galaxies have secure detections of both
H$\alpha$ and H$\beta$ meaning we are not forced to reject any DYNAMO
galaxy from this analysis. We
calculate $E(B-V)_{gas}$ using the equation \citep{calz00}: 
\begin{equation}\label{equation:hahb}
  E(B-V)_{gas}=\frac{log(R_{obs}/R_{int})}{0.4[k(\lambda_{H\alpha})-k(\lambda_{H\beta})]}
\end{equation}
where $R_{obs}$ and $R_{int}$ are the observed and intrinsic H$\alpha$
to H$\beta$ ratios \citep[assuming $R_{int}$=2.86 from Case B
recombination,][]{osterbrock89} and $k(\lambda)$ is the attenuation curve normalised such that
$k(B)-k(V)=1$. Here we employ $k(\lambda)$ from \citet{card89},
which is appropriate for star-forming regions and is commonly used for
correcting emission line fluxes at low and high redshifts. 

\subsubsection{DYNAMO $E(B-V)_{stars}$}

$E(B-V)_{stars}$ is
computed from $\tau_{V}$ taken from the MPA-JHU VAC. \citet{kauffmann03a} compute
synthetic $g$, $r$, and $i$ magnitudes from SDSS spectra and 
compare these to similar, synthetic $g-r$
and $r-i$ colours from a library of single stellar population
spectra constructed according to the models of
\citet{bruzual03}. Through this process, they obtain an estimate of
$\tau_{V}$ affecting the emitted starlight. We convert
this to $E(B-V)_{stars}$ using the equation \citep{calz01}:
\begin{equation}
  \tau_{V} = 0.921 \times E(B-V)_{stars} \times k(\lambda)
\end{equation}
In this case we assume an attenuation curve with a
functional form of $k(\lambda) \propto \lambda^{-0.7}$ following
\citet{charlot00}, which is the same procedure used by
\citet{kauffmann03a}. 

As galaxies move to higher redshift, optical photometric bands cover a
smaller range in rest frame wavelength, and this must be taken into
account using a K-correction. Spectral models used by \citet{kauffmann03a}
include redshift and thus provide restframe B and V magnitudes, which
should roughly account for the K-correction. Furthermore, at redshifts
covered by the DYNAMO sample, \citet{westra10} show empirically that
K-corrections for the $g$, $r$, and $i$ bands are similar at around
0-0.4 mag. Thus the $g-r$ and $r-i$ may change by a maximum of
0.4 mag due to the effects of spectral
broadening below $z = 0.15$. \citet{westra10} also find
that that galaxies where stellar emission is dominated by young
stellar populations, as is the case for many DYNAMO galaxies
\citep{green14,bassett14}, require the smallest K-correction, thus the
difference in observed colours a K-correction would induce is likely to be less
than 0.1 mag. For these
reasons, such a correction would have a negligible effect on
our results.

\subsection{DYNAMO Integrated Attenuation Results}

\subsubsection{$E(B-V)_{gas}$ vs $M_{*}$ and SFR}

\begin{figure}
  \includegraphics[width=3.6in]{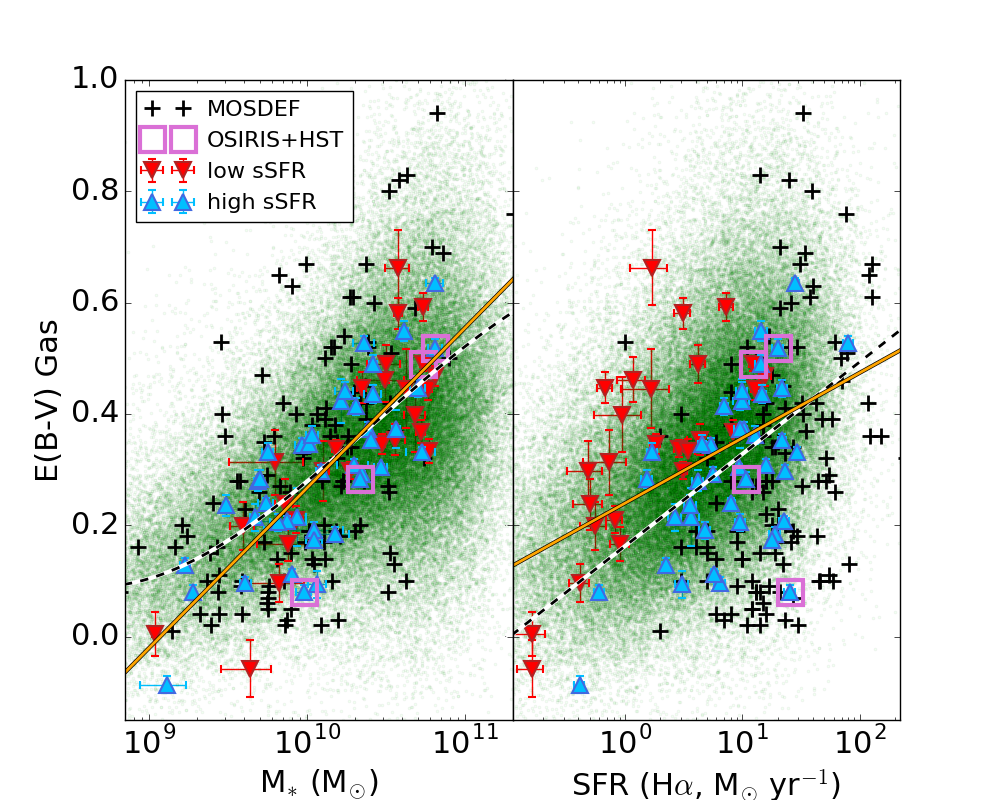}
  \vspace{-2em}
  \caption{The relationships between $E(B-V)_{gas}$ measured from the
    Balmer decrement and stellar mass (left panel) and $E(B-V)_{gas}$
    and SFR (H$\alpha$, right panel). DYNAMO galaxies with sSFR larger
    than the median sSFR for DYNAMO are plotted with blue upward
    triangles, and galaxies with lower sSFR's are plotted with red
    downward triangles. Overplotted in each panel are
    the low redshift relationships between mass/SFR and $E(B-V)_{gas}$ from the
    SDSS \citep[black dashed line,][]{garn10b}, SDSS datapoints in green (taken from the
    MPA-JHU VAC). Note that the apparent difference between the fit
    and SDSS points is a visual effect due to the density of the data
    points. We also plot data for high redshift galaxies taken from the MOSDEF
    survey taken from \citet{reddy15}. We also plot linear fits to
    log$_{10}$($M_{*}$) and log$_{10}$(SFR) for DYNAMO galaxies as
    solid orange lines in each panel, which show DYNAMO galaxies to be
    in rough agreement with \citet{garn10b}. The MOSDEF sample also
    roughly follows the SDSS trend with $M_{*}$ \citep[considering the
    large scatter found by][]{garn10b} while following a steeper
    relation with SFR mirrored by high sSFR DYNAMO galaxies}\label{figure:ch4agascorr}
\end{figure}

\citet{garn10b} find attenuation of ionized gas to
correlate with both $M_{*}$ and SFR in a sample of star-forming
SDSS
galaxies (with AGN excluded). In Figure \ref{figure:ch4agascorr} there
is an apparent offset between the fit of \citet{garn10b} and the SDSS
points, however this is simply a visual effect due to the density of
plotted points. From this sample \citet{garn10b} perform fits to the 
$A_{H\alpha}$ vs SFR and $A_{H\alpha}$ vs $M_{*}$ relationships
independently. We compare the results for DYNAMO and MOSDEF galaxies to these local relations in Figure
\ref{figure:ch4agascorr} where the results of \citet{garn10b} are
shown as black dashed lines. These relations represent polynomial fits
of the form:
\begin{equation}\label{equation:garn10b1}
  A_{H\alpha} = 0.53 + 0.54
  \textrm{log}_{10}(\textrm{SFR/M}_{\odot}\textrm{ yr}^{-1})
\end{equation}
\begin{equation}\label{equation:garn10b2}
  A_{H\alpha} = 0.91 + 0.77 M' + 0.11 M'^{2} - 0.09 M'^{3}
\end{equation}
where $M'$ = log$_{10}$($M_{*}$/10$^{10}$
M$_{\odot}$).

In general, \citet{garn10b} find that attenuation of line emitting gas
increases with both SFR and $M_{*}$, as well as metallicity, in non-active, star-forming SDSS
galaxies. The authors conclude that galaxy mass is the best indicator
of attenuation and that, considering $M_{*}$,
Equation \ref{equation:garn10b2} is able to predict $A_{H\alpha}$ with
an error
of 0.28 mag. They suggest that this is because as a galaxy increases
in $M_{*}$ via star-formation it also continuously builds up a dust
reservoir thus gradually increasing in $A_{H\alpha}$. More massive
galaxies form greater amounts of stars during bursts and they retain a
larger fraction of metals produced through star
formation resulting in $A_{H\alpha}$ vs SFR and $A_{H\alpha}$ vs
metallicity relationships, which are secondary to the dependence on
$M_{*}$. This result is in rough agreement with a number of other
similar studies \citep[e.g.][]{brinchmann04,dacunha10a,cortese12}.

In
Figure \ref{figure:ch4agascorr}, we check for similar correlations in
DYNAMO and high redshift MOSDEF data. Overplotted as a dashed line in
each panel are the
polynomial fits from SDSS \citep{garn10b} that relates $A_{H\alpha}$ to
$M_{*}$ and SFR (equations \ref{equation:garn10b1} and
\ref{equation:garn10b2}) where we have
converted $A_{H\alpha}$ to $E(B-V)_{gas}$
assuming a \citet{calz00} attenuation curve with $R_{V}=4.0$. These
fits are shown as black dashed lines. We have
also plotted in green SDSS galaxies identified as star-forming,
non-AGN with signal to noise cuts for H$\alpha$ and H$\beta$ of 20 and
3 respectively, which matches the requirements of \citet{garn10b}. 
Finally, we plot a linear fits to all DYNAMO points as orange solid lines.

Considering the left panel of Figure \ref{figure:ch4agascorr}, we find
a good agreement between the SDSS
$M_{*}$-$E(B-V)_{gas}$ relation and the DYNAMO sample. MOSDEF galaxies
roughly follow these trends as well, in agreement with the conclusion
from \citet{garn10b} that $M_{*}$ is a fundamental driver of the dust
content of galaxies.

In the right panel of Figure \ref{figure:ch4agascorr}, we
find that correlations between $E(B-V)_{gas}$ and SFR are less
clear. Considering the entire sample of DYNAMO galaxies we find a weak
correlation between $E(B-V)_{gas}$ and SFR (Kendall $\tau$ = 0.31,
3.73$\sigma$ significance). In the right panel of Figure
\ref{figure:ch4agascorr} we again plot a linear fit
to DYNAMO points as an orange solid line, finding this to be comparable to SDSS
relation of \citet{garn10b}. We note however that for both
SDSS and DYNAMO galaxies there is significant scatter about the fitted
relationships. 

The relationship between SFR vs $E(B-V)_{gas}$ for MOSDEF
galaxies appears to follow a steeper relationship compared
to that of SDSS and DYNAMO.  We separate
DYNAMO galaxies into low and high specific SFR (sSFR=SFR/$M_{*}$) with the cut at
log$_{10}$(sSFR)=5.35$\times$10$^{-10}$ yr$^{-1}$, the median of the DYNAMO
sample. We find
that at high sSFR galaxies appear to follow a similar trend between
SFR and $E(B-V)_{gas}$ to that found for
the MOSDEF survey while low sSFR
galaxies remain uncorrelated. This may be due to an intrinsically
different relationship for galaxies with very high sSFR. This may also reflect
the difficulty in obtaining reliable H$\alpha$ \textit{and} H$\beta$ 
fluxes (which are
both necessary to measure $E(B-V)_{gas}$) for galaxies with a large
$E(B-V)_{gas}$ and low SFR at high 
redshift. 

We note here that \citet{garn10a} find no evidence of an
evolution in the SFR vs $E(B-V)_{gas}$ relation compared to SDSS up to $z=0.84$
considering observations of star-forming galaxies from the High
Redshift Emission Line Survey \citep[HiZELS,][]{geach08}.
\citet{dominguez13} further extend this to $z\sim1.5$ using Wide Field Camera
3 (WFC3) IR Spectroscopic Parallel (WISP) Survey data. The lack of
overlap between the MOSDEF sample and low 
sSFR DYNAMO galaxies may result from incompleteness of the MOSDEF sample
with regards to H$\beta$ detections. \citet{garn10b} note that by
including galaxies with low H$\alpha$ S/N in their sample, the median
$A_{H\alpha}$ increases by 0.35 mag, which is equivalent to a
change in $E(B-V)_{gas}$ of $\sim$0.1. This is comparable to the average
difference between low and high sSFR DYNAMO galaxies at fixed SFR,
meaning that incompleteness in the MOSDEF survey with regards to low
H$\alpha$ flux galaxies may account for the lack of galaxies with low
SFR and high $E(B-V)_{gas}$ in their sample.

\subsubsection{$E(B-V)_{gas}$ vs $E(B-V)_{stars}$}

\begin{figure}
  \includegraphics[width=3.6in]{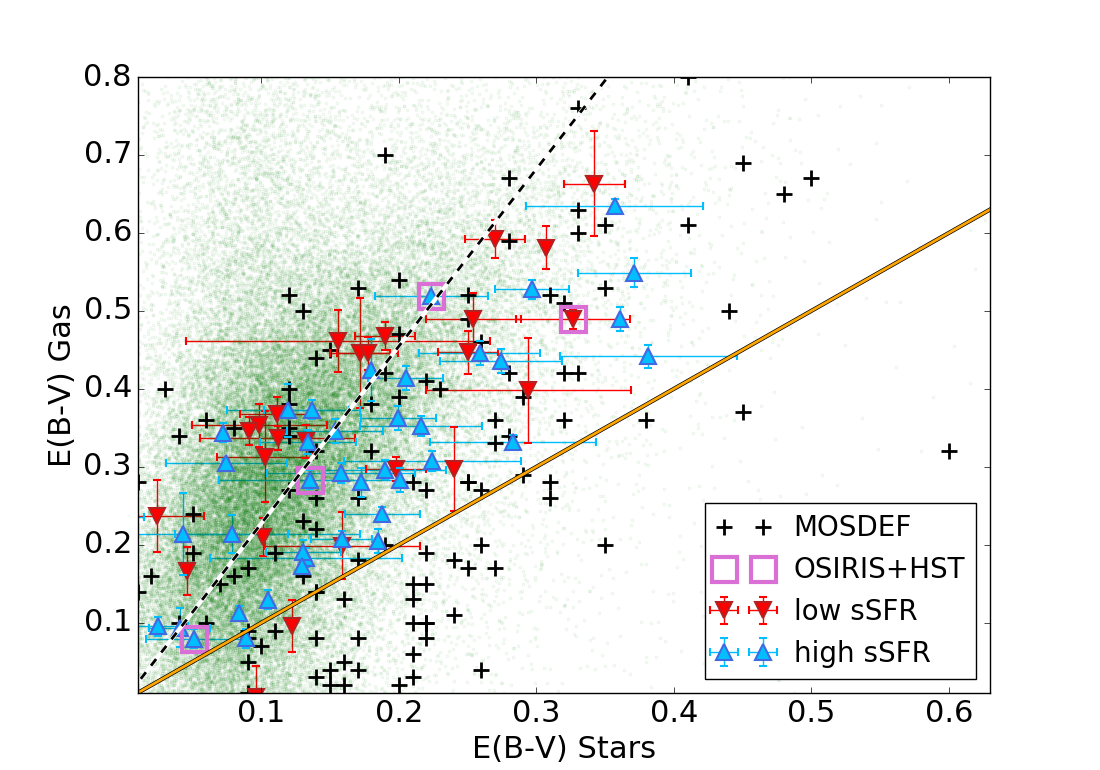}
  \vspace{-2em}
  \caption{Here we compare integrated measures of $E(B-V)_{gas}$ from the Balmer decrement
    with measurements of $E(B-V)_{stars}$. Low sSFR DYNAMO galaxies are given
    by red downward triangles, high sSFR DYNAMO galaxies by blue
    upward triangles where the separation is at the median sSFR of the
    DYNAMO sample. High redshift MOSDEF galaxies are given by
    black crosses. These datapoints are plotted over the ~55,000 SDSS data release 4 galaxies from the MPA-JHU
  VAC in green, the parent sample of the DYNAMO survey. We also indicate the
  local relation $E(B-V)_{stars} = 0.44 \times E(B-V)_{gas}$
  \citep{calz97} as a black dashed line, and the 1-to-1 relation
  using a solid line. Galaxies for which resolved attenuations are
  investigated are shown using purple open
  squares.}\label{figure:ch4sedahacomp}
  \vspace{-.2in}
\end{figure}

We also investigate the relation found for local starbursts
$E(B-V)_{stars}=0.44 \times E(B-V)_{gas}$ \citep{calz97} for $z\sim0.1$
galaxies with SFRs more typical of the $z>1$ universe. This relation has come under
recent scrutiny, particularly at high redshifts, and its applicability
to turbulent high redshift disks is unclear
\citep{price14,reddy15}. In Figure \ref{figure:ch4sedahacomp} we plot
$E(B-V)_{stars}$ versus $E(B-V)_{gas}$ for DYNAMO and MOSDEF
galaxies. 
The solid line in this plot gives the 1-to-1 relation while the dashed line
shows the local relation, $E(B-V)_{stars}=0.44\times E(B-V)_{gas}$. 
Again, we separate low and high sSFR DYNAMO galaxies as in Figure
\ref{figure:ch4agascorr}. DYNAMO and MOSDEF galaxies are plotted over the
values for SDSS star-forming, non-AGN in green, taken from the MPA-JHU VAC.

We find good agreement in the relationship between $E(B-V)_{stars}$
and $E(B-V)_{gas}$ when comparing the DYNAMO and MOSDEF
samples. Many galaxies from both surveys appear
to be well described by the local relation, although a significant
portion of both samples fall below this relation, closer to the 1-to-1
line. We perform linear fits to both datasets using the IDL routine
poly\_fit finding $E(B-V)_{stars}=0.78 \pm 0.08 \times E(B-V)_{gas}$
and $E(B-V)_{stars}=1.07\pm0.25 \times E(B-V)_{gas}$ for DYNAMO and
MOSDEF galaxies, respectively. The larger scatter exhibited by MOSDEF
galaxies in Figure \ref{figure:ch4sedahacomp} is reflected in the
larger uncertainty in the fitted slope. In particular, low
$E(B-V)_{stars}$ MOSDEF galaxies scatter to very low $E(B-V)_{gas}$
(below the 1-to-1 line) thus driving the correlation very close to the
1-to-1 relation. At higher $E(B-V)_{stars}$, MOSDEF and DYNAMO
galaxies appear to be in better agreement. 

Overall, these results
suggest that applying the local relation,
$E(B-V)_{stars}=0.44 \times E(B-V)_{gas}$, only occasionally provides a reasonable
estimate of the attenuation suffered by ionized gas emission where
reliable direct estimates of $E(B-V)_{gas}$ (e.g. the Balmer
decrement) are not available. In some cases, particularly for more
highly attenuated galaxies, this relationship may lead to an
overestimate of $E(B-V)_{gas}$, and in turn, an overestimate of SFR if
this is measured from $F($H$\alpha)$. We also note that three of four
galaxies making up our resolved
attenuations sample fall very close to the $E(B-V)_{stars}=0.44 \times
E(B-V)_{gas}$ line, while one, galaxy D 13-5, falls roughly halfway
between this relation and the 1-to-1 line. These differences in
$E(B-V)_{stars}$ vs $E(B-V)_{gas}$ may be related to dust geometry in
these galaxies, an effect that can not easily be accounted for in
integrated measurements. We investigate this possibility in the remainder of this
work using our combined HST and OSIRIS observations of H$\alpha$ and
Pa$\alpha$.

\section{Resolved Attenuation on Sub-kpc Scales}\label{section:ch4resavmeas}

\subsection{Measuring Resolved Attenuations}

\subsubsection{IFS Emission Line Maps}\label{section:ch4ifselm}

Prior to describing our method of producing IFS emission line maps, we wish to be
explicit about terms. The term ``spaxels'' refers to the individual
spectra associated with each 2D resolution element of a 3D datacube,
thus distinguishing them from the ``pixels'' of a more traditional
photometric detector. 

We produce maps of emission lines from our IFS datacubes following a similar
procedure to that outlined in \citet{bassett14}.  Briefly, we fit a
single Gaussian profile to the emission line in question with a
linear fit to the surrounding continuum included. The total flux is
taken as the sum of the continuum subtracted spectrum at all positions where the fitted
gaussian profile deviates from the continuum by more than 0.1\%. This
procedure is performed in each spaxel individually. We mask pixels in the final
flux maps in which the signal to noise of a given emission line is
less than one. We also mask pixels that may be
associated with noise spikes (e.g. residual cosmic rays) by performing
cuts on the measured velocity and velocity dispersion (also taken from
our gaussian fits). Excluded are fits where velocities deviate by > 250 km s$^{-1}$
from the systemic velocity or where measured velocity dispersions less than the
instrumental resolution measured from our arc exposures. Maps of
Pa$\alpha$ are shown in comparison to matched HST images in Section
\ref{section:ch4resavres}. For galaxy G 20-2, we use the same
procedure to map both [OIII] (5007 {\AA}) and H$\beta$ allowing us to produce maps of
[OIII]/H$\beta$. This is discussed in more detail in Section \ref{section:ch4resavres}.

\subsubsection{Matching H$\alpha$ to Pa$\alpha$ Data}\label{section:ch4psfmatch}

A key to accurately measuring variations in attenuation in individual
DYNAMO galaxies will be reliably matching the PSF of
HST ($\sim$0$\farcs$1 resolution) to that of OSIRIS ($\sim$0$\farcs$6
resolution wings; $\sim$0$\farcs$15 resolution core).
The overall difference in the resolution achieved between the two
datasets is small due to the use of
AO in our OSIRIS observations, which significantly reduces the core
size of the PSF. One typical feature of AO systems, however, is that a
large fraction (often a majority, $\sim$60\% in our observations) of
the light from a point source 
will be distributed in a broad wing extending significantly further
than the PSF core. Because the HST PSF is more compact (and contains a
larger fraction of the light in its core), measurements of the
Pa$\alpha$ to H$\alpha$ ratio using OSIRIS and HST data will be
artificially high in clump cores and artificially low in non-clump
regions if the HST data is not matched to the AO PSF of OSIRIS.

We employ the IRAF task \textit{psfmatch} to properly account for
differences between the HST and OSIRIS PSFs. This task creates a convolution
kernel, which can transform an input PSF to into a given output PSF
through a comparison of the two in Fourier space. For our HST data, we
created a PSF by averaging three stars that fall within our observed
region for each galaxy, normalized to their peak flux value. Stars are
selected with large angular distances from other sources to
minimize contamination in the outskirts of our empirical PSF. We perform a similar
procedure to create an empirical OSIRIS PSF using continuum averaged
IFS observations of tip-tilt stars during a single night. 

The initial PSF size of both our HST
observations and the core size of the OSIRIS PSF are comparable at
around $\sim$0$\farcs$15. This means our PSF matching procedure,
based on Fourier transformations of our PSF images, tends to
over-correct our HST photometry. Due to this, we choose to also PSF
match our
OSIRIS imaging to the raw PSF of HST as well using the same
empirically produced PSFs described above. This second PSF matching
step results in a very slight difference in our final Pa$\alpha$ image
as we are matching to a slightly narrower PSF. We note that our PSF
matching procedure, if
done improperly could introduce trends in our maps of attenuation for
our sample and furthermore the wings of the AO PSF can introduce
trends even after proper matching has been performed. In Appendix
\ref{section:ch4psfmtest}, we perform tests in
order to address this, finding this not to be a large factor in our
analysis. 

We employ the IRAF tasks \textit{geomap} and
\textit{gregister} to register our HST images to our OSIRIS maps. As
an input, \textit{geomap} requires the pixel coordinates of a
number 
of reference positions for both images. Due to the irregular, clumpy
morphologies of galaxies in our sample, we typically observe 4-5 marginally
resolved star-forming regions that provide ideal reference points for
this process. A caveat to this is that we assume H$\alpha$ and
Pa$\alpha$ peaks to be coincident, however we consider this to be
reasonable based on visual inspection of the two images. We estimate the centres of clumps from both datasets
using the IRAF task \textit{ellipse} and use these values to create input files
for \textit{geomap}. The \textit{geomap} task outputs a database containing
the appropriate coordinate transformations that are used as input for
\textit{gregister}. Together, these tasks provide a flexible tool for image
matching that simultaneously handles reflection, rotation,
translation, and magnification. The final results of our PSF matching
and image registration are presented in Section
\ref{section:ch4resavres}. 

\subsubsection{Correcting $F($H$\alpha$) based on SPIRAL/WiFeS [NII]/H$\alpha$}\label{section:ch4NII_Ha}

We correct our HST H$\alpha$ photometry for contamination from two [NII] emission
lines (at 6548 {\AA} and 6583 {\AA}) as these are known to fall within the spectral window of our
ramp filter. To accomplish this, we investigate the contribution to
our narrow band images from these lines using the original DYNAMO observations taken with the
SPIRAL and WiFeS IFS. These observations are often marginally resolved
spatially but can provide H$\alpha$ and [NII] fluxes and any spatial variations on 2-5
kpc scales.

In each spaxel of our SPIRAL and WiFeS datacubes we measure the ratio
of the H$\alpha$ line flux to the total continuum subtracted flux in
the HST
ramp filter with the central wavelength matched to those of our
observations of each galaxy. Here we have used the functional form for
the HST ramp filter provided on the Space Telescope Science Institute
website. Spaxel-to-spaxel H$\alpha$ fluxes are measured by fitting
a gaussian profile giving the H$\alpha$ flux,
F(H$\alpha$). We then make an estimate of the total flux
of all emission lines (H$\alpha$ plus the [NII] doublet) in our narrow
band filter, F(HST$_{NB}$). This is done by subtracting a linear
continuum fit to the spectrum in each spaxel and summing the total
flux contained in the HST ramp filter. The ratio of the fluxes measured in this way, F(H$\alpha$)/F(HST$_{NB}$), represents the
multiplicative factor needed to corrected our HST H$\alpha$ photometry
for the presence of [NII] in our HST ramp filter. In our three disk
galaxies, D 13-5, G 04-1 and G 20-2, we find that the maps of
F(H$\alpha$)/F(HST$_{NB}$) exhibit central values of 0.4-0.6 with a
sudden increase to $\sim$0.9 at the large radii where H$\alpha$
uncertainties are the highest. We also note that, although there is
known variation in the spectral coverage of the ACS ramp filters over
the HST field of view, observed galaxies are significantly smaller
than the HST field of view and, similarly, the width of the spectral
features of interest are small compared to the width of the ramp filter
transmission curve.
For these reasons we choose to apply a flat correction for the
presence of [NII] emission in our HST ramp filter based on
integrated measurements of our SPIRAL datacubes.

This correction is achieved by
summing the spaxels within two r-band $r_{e}$ of the galaxy centre, creating a characteristic
spectrum for each galaxy. We measure F(H$\alpha$)/F(HST$_{NB}$) as
described above on this summed spectrum giving the global correction
value that is applied to all spaxels. 
The form of our correction is given as:
\begin{equation}\label{equation:ch4niihaeq}
  F(H\alpha)=F_{HST}\left(\frac{F(H\alpha)}{F(HST_{NB})}\right)
\end{equation}
Where $F_{HST}$ is the total flux of our HST narrowband observations.
This correction is applied to each pixel producing a corrected
H$\alpha$ image. Our correction factors for H$\alpha$ for each galaxy
are presented in Table \ref{table:ch4fcorr}. Galaxy H 10-2 has a
significantly higher correction factor than the other three, and this
is simply due to this galaxy having very weak $[NII]$ emission
relative to H$\alpha$. 

\begin{table}
  \caption{H$\alpha$ and Pa$\alpha$ Flux Correction Factors}
  \centering
  \begin{tabular}{ c c c }
  \hline\hline
  Galaxy & $F(H\alpha)$/$F(HST_{NB})$ correction & $c_{Pa\alpha}$ correction\\
    & H$\alpha$, Section \ref{section:ch4NII_Ha} & Pa$\alpha$, Section \ref{section:ch4bdcorr}\\
  \hline
  D 13-5 & 0.45 & 2.35\\
  G 04-1 & 0.40 & 0.77\\
  G 20-2 & 0.46 & 0.94\\
  H 10-2 & 0.60 & 0.76\\
  \hline 
  
  \end{tabular}
\label{table:ch4fcorr}
\end{table}

\subsubsection{Correcting $F($Pa$\alpha$) Based on SDSS Balmer Decrement}\label{section:ch4bdcorr}

As mentioned in Section \ref{section:ch4OSIRISd} the
spectrophotometric flux calibration of our OSIRIS datacubes suffers
from a $\sim$30\% systematic uncertainty due to variations in the
Strehl ratio (the ratio of the maximum observed intensity to that of
an ideal, diffraction limited system) of the OSIRIS PSF during a given night of
observations. For this reason we choose to perform a correction to our
Pa$\alpha$ maps by comparing integrated Pa$\alpha$/H$\alpha$ values
based on the data presented here to H$\alpha$/H$\beta$ values measured
from SDSS taken from the MPA-JHU Value Added Catalog \citep{tremonti04}. 

For this correction, we take the integrated $E(B-V)_{gas}$ measured in
Section \ref{section:intavdata} from the Balmer Decrement using
Equation \ref{equation:hahb}.
We assume the value calculated in this way to be the characteristic global
value of $E(B-V)_{gas}$ for a given galaxy as the emission lines in the SDSS
spectra are more likely internally consistent compared with our
independent observations of H$\alpha$ and Pa$\alpha$.
Using this value, we can now predict the value of
Pa$\alpha$/H$\alpha$ that would be detected in a region matching the
SDSS fibre observation for each galaxy. This is done using Equation \ref{equation:hahb}
modified in consideration of the Pa$\alpha$ and H$\alpha$ emission
lines:
\begin{equation}\label{equation:paha}
  E(B-V)_{gas}=\frac{log(R_{obs}/R_{int})}{0.4[k(\lambda_{Pa\alpha})-k(\lambda_{H\alpha})]}
\end{equation}
We insert the $E(B-V)_{gas}$ value calculated from the SDSS spectrum
and again assume a \citet{card89} $k(\lambda)$ and Case B
recombination, which gives an intrinsic ratio of Pa$\alpha$ to
H$\alpha$ of 0.123. We then solve Equation \ref{equation:paha} for
$R_{obs}$, giving the expected observed integrated ratio of Pa$\alpha$
to H$\alpha$. 

\begin{figure*}
  \centering
  \includegraphics[width=.8\textwidth]{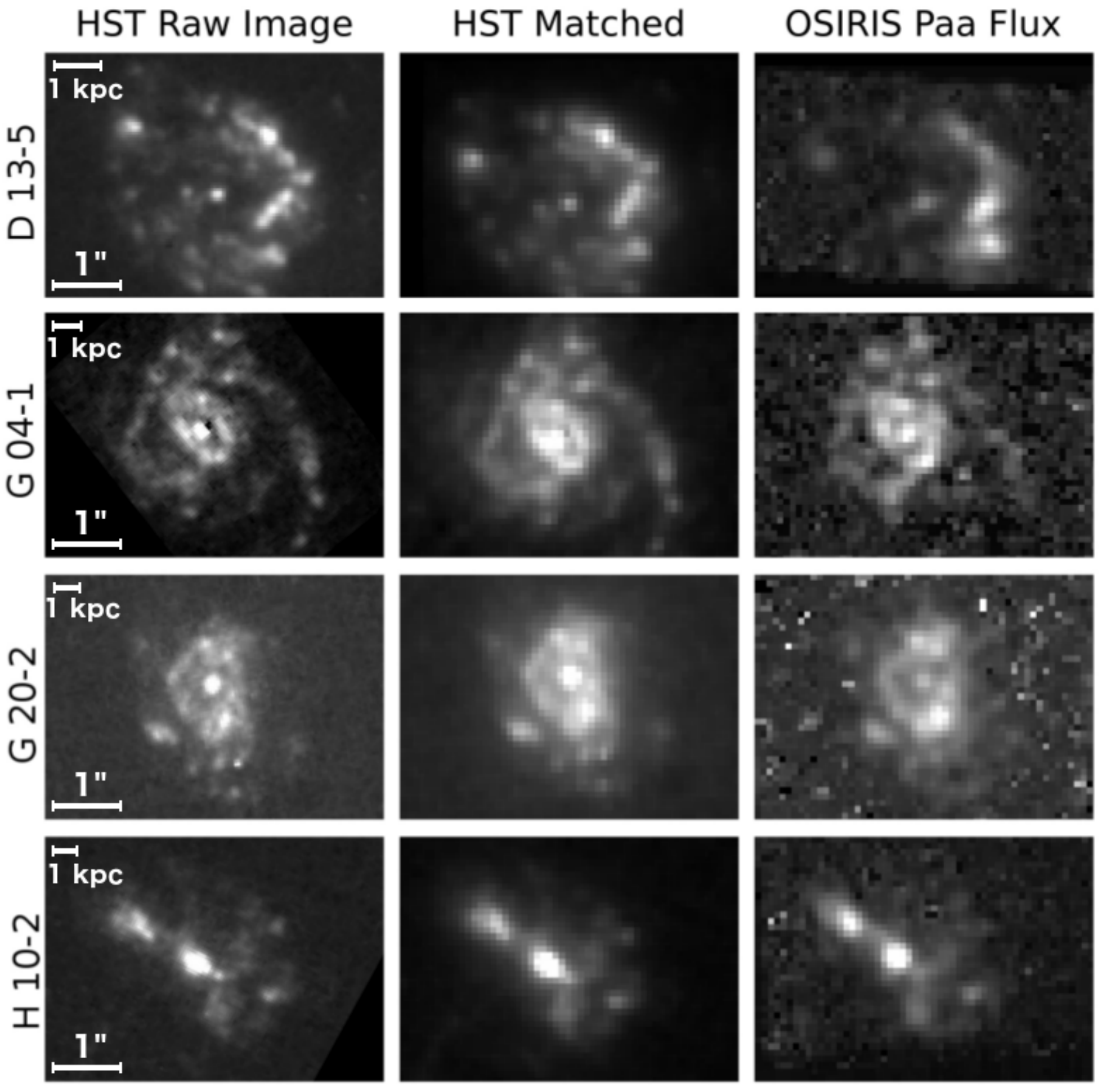}
  \caption{Our PSF matching and registration procedure applied to
    each of the four galaxies in our combined HST (H$\alpha$) and Keck
    OSIRIS (Pa$\alpha$) sample with
    each galaxy shown in a single row. In the left panel of each row
    we also include scale bars in the bottom and top left corners
    scale bars indicating a size of 1$\farcs$0 and 1 kpc,
    respectively. The colour scaling here is arbitrary
    as we simply wish to illustrate the results of PSF matching and
    image registration. There is not a large difference between the
    raw and matched HST photometry due to the core of the Keck AO PSF
    being only slightly larger than the PSF of HST. The most
    noticeable effects are a slight blurring due to the broad wing of
    the AO PSF as well as a reduction of the pixel scale to match that
    of OSIRIS. We note the remarkable agreement in visual morphology
    between the two datasets.}\label{figure:ch4psfmm}
\end{figure*}

We next measure the integrated Pa$\alpha$/H$\alpha$ as observed for
our galaxies by HST and OSIRIS in the following
way. We first sum the flux of our PSF-matched HST
H$\alpha$ photometry in an artificial aperture matched to that of
the SDSS spectroscopic fibre. Similarly, we create an integrated spectrum of our OSIRIS data by
summing all spaxels contained within an artificial SDSS aperture. The
H$\alpha$ flux is calculated from the HST counts and the Pa$\alpha$
line flux is measured from our integrated spectrum as described in
Section \ref{section:ch4ifselm}. This measurement of Pa$\alpha$/H$\alpha$ is
thus taken in  a geometric region comparable to that sampled by the
SDSS fibre. In this way we can provide a correction for each galaxy by
comparing our observed value with the expected
$R_{obs}$ from SDSS computed above.

As H$\alpha$ fluxes from HST photometry are far more reliable than
Pa$\alpha$ fluxes from our AO assisted IFS observations, we choose to
correct the latter while assuming the former to be correct. The
multiplicative correction factors, $c_{Pa\alpha}$, for each of our Pa$\alpha$
observations are calculated by dividing the expected value of
$R_{obs}$ by the value we observe. By performing this
correction we correct our Pa$\alpha$ fluxes based on a more robust
measurement of the overall attenuation in our system, which will
provide more reliable \textit{relative} measurements (in a spatially resolved sense) as we
show in Section \ref{section:ch4resavres}. Values of $c_{Pa\alpha}$ are quoted in
Table \ref{table:ch4fcorr} for each galaxy. 

\begin{figure*}
  \centering
  \includegraphics[width=\textwidth]{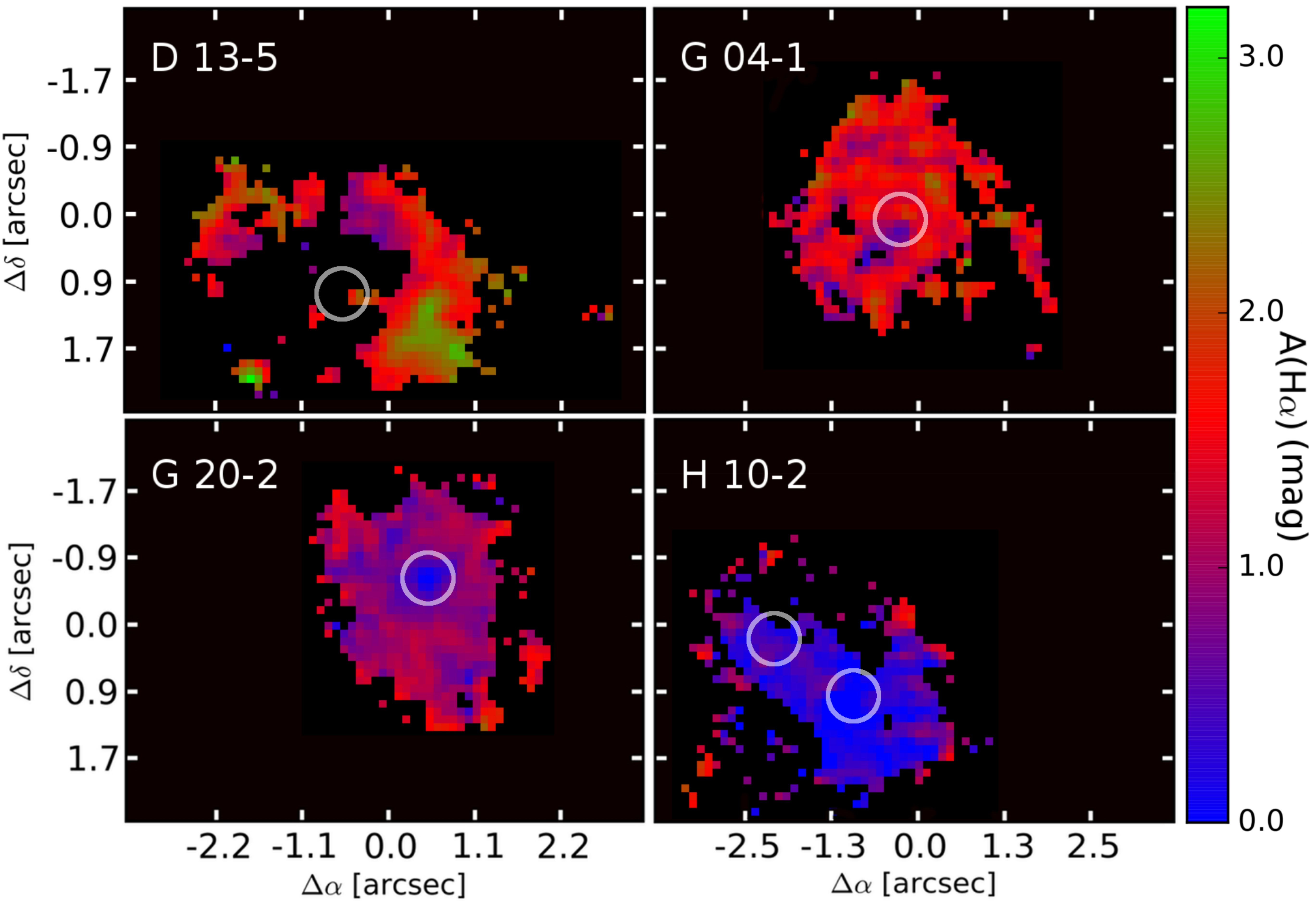}
  \caption{Maps of $A_{H\alpha}$ produced from the ratio of our
    Pa$\alpha$ and H$\alpha$ maps shown in Figure
    \ref{figure:ch4psfmm}. The spread of values in individual galaxies
  are consistent with attenuations typical of local star-forming
  galaxies. Approximate locations of the continuum peaks in each
  galaxy are indicated with white circles. See text for detailed descriptions of individual galaxies.}\label{figure:dustmaps}
\end{figure*}

Galaxies G 04-1, G 20-2,
and H 10-2 have corrections between $\sim$10\% and 30\% while the
correction for D 13-5 is significantly larger. As we show in the next
Section, the distribution of dust in D 13-5 appears to be clumpier
than that of the other three galaxies, and such a clumpy dust
distribution may result in an intrinsically steeper attenuation law
\citep[e.g][]{inoue05}. By assuming a shallow attenuation law here, we may
underestimate the difference between Pa$\alpha$ and
H$\alpha$, thus overestimating the expected observed ratio
Pa$\alpha$/H$\alpha$ and, in turn, $c_{Pa\alpha}$ for D
13-5. Regardless, we adopt a shallow attenuation law for consistency
and again stress that, while the absolute value of $A_{H\alpha}$ may
be systematically offset, the \textit{relative} attenuations across
the face of this galaxy are still relevant. 

\subsubsection{Mapping $A_{H\alpha}$}

Having spatially matched our H$\alpha$ photometry to our Pa$\alpha$
IFS flux maps and applied flux corrections to both sets of
observations, we can then produce resolved maps of attenuation of the four observed
galaxies. First we simply produce Pa$\alpha$/H$\alpha$ maps by
dividing our OSIRIS Pa$\alpha$ flux maps by our HST H$\alpha$
photometry. Next we convert the value in each pixel to $E(B-V)_{gas}$
using Equation \ref{equation:paha}, which assumes Case B
recombination. Finally we convert $E(B-V)_{gas}$ to $A_{H\alpha}$
using a \citet{card89} extinction curve and assuming $R_{V}$ =
3.1. The resulting maps of $A_{H\alpha}$ for each of our galaxies are
presented in the following Section.

\subsection{Results: Resolved Attenuation of Four Clumpy Galaxies}\label{section:ch4resavres}

\subsubsection{Maps of Attenuation From R(Pa$\alpha$,
  H$\alpha$)}\label{section:ch4dustmaps}

\begin{figure}
  \centering
  \includegraphics[width=\columnwidth]{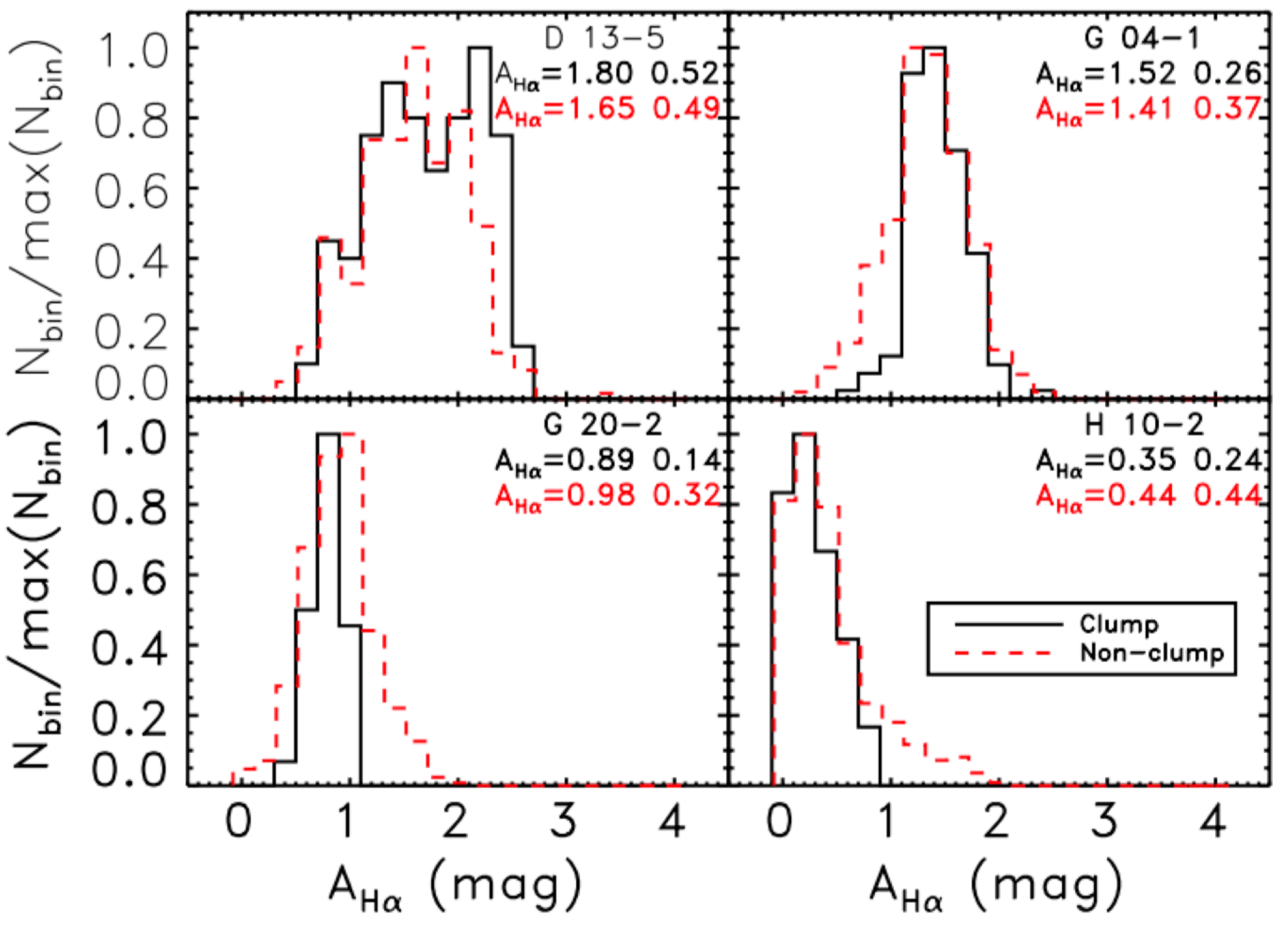}
  \caption{Histograms of $A_{H\alpha}$ in clump versus non-clump spaxels in each of our
  observed galaxies where clump spaxels are defined in Section
  \ref{section:ch4clmeas}. Within a given galaxy, the average attenuation in clumpy and
  non-clumpy regions are quite similar. Based on a two sample KS-test
  we reject the hypothesis that $A_{H\alpha}$ values come from the
  same parent distribution. However, this is most likely due to the
  fact the largest $A_{H\alpha}$ values are found in 
  non-clump spaxels, resulting from less reliable
  Pa$\alpha$ flux measurements outside of bright clumps. In each panel
the mean and standard deviation of clump and non-clump $A_{H\alpha}$
values are indicated with text colours matching the corresponding histograms.}\label{figure:dusthists}
\end{figure}

PSF-matched and registered H$\alpha$ and Pa$\alpha$ maps of
each galaxy are shown in Figure \ref{figure:ch4psfmm}, as well as the unmatched
H$\alpha$ images for reference. The matched HST and OSIRIS maps show a
remarkable agreement in visual morphology suggesting an absence of
patchy dust. In particular the locations and relative brightness of
clumps, the term used to describe individual H$\alpha$ and Pa$\alpha$
peaks \citep[see also,][for more on our definition of ``clumps'']{fisher16}, are well matched between the two datasets. This is consistent with the appearance of 
$A_{H\alpha}$ maps shown in Figure \ref{figure:dustmaps}. In general,
we find overall attenuation values consistent with
values found for local star-forming samples or roughly 0.0 mag $\leq$
$A_{H\alpha}$ $\leq$ 3.0 mags. Within an individual galaxy variations of $A_{H\alpha}$ are
typically less than 1.0 mag. This is also shown as histograms of
$A_{H\alpha}$ in individual spaxels in Figure
\ref{figure:dusthists}.

Overall we find no general
correlation between the locations of clumps and increased $A_{H\alpha}$. In galaxy D 13-5 and in
the central ring of G 04-1 we find variation in $A_{H\alpha}$ of
$\sim$0.5-3 mag, but regions of increased attenuation do not appear
coincident with any one particular clump. This could indicate a clumpy
distribution of dust in these galaxies that is not spatially
correlated with the locations of star-forming clumps. We also note
that, due to the lower redshift of D 13-5, the OSIRIS field of view of
this galaxy covers only the central star-forming region meaning a
comparison of the full OSIRIS observation of D 13-5 and the central
region of G 04-1 is apropos. 

We find no evidence of highly
attenuated regions in any of our galaxies comparable to regions seen
in low redshift ULIRGS \citep[e.g.][]{piqueras13}. We also
observe H$\alpha$ emission covering the entire
central regions of these four galaxies strongly implying that we
observe no optically thick regions such as those expected for high
redshift SMGs \citep{blain02}. We can say with confidence
that DYNAMO galaxies presented here do not contain highly attenuated
clumps hosting any appreciable star-formation. Clumps such as these would be
readily apparent in our OSIRIS maps of Pa$\alpha$ as well as our maps
of $A_{H\alpha}$. For our small sample, clumps
observed at H$\alpha$ represent the full census of
highly star-forming regions. The applicability of this observation to
clumpy galaxies at high redshift is still unclear, however. We now
provide a more detailed description of each galaxy individually.

\begin{figure}
  \includegraphics[width=\columnwidth]{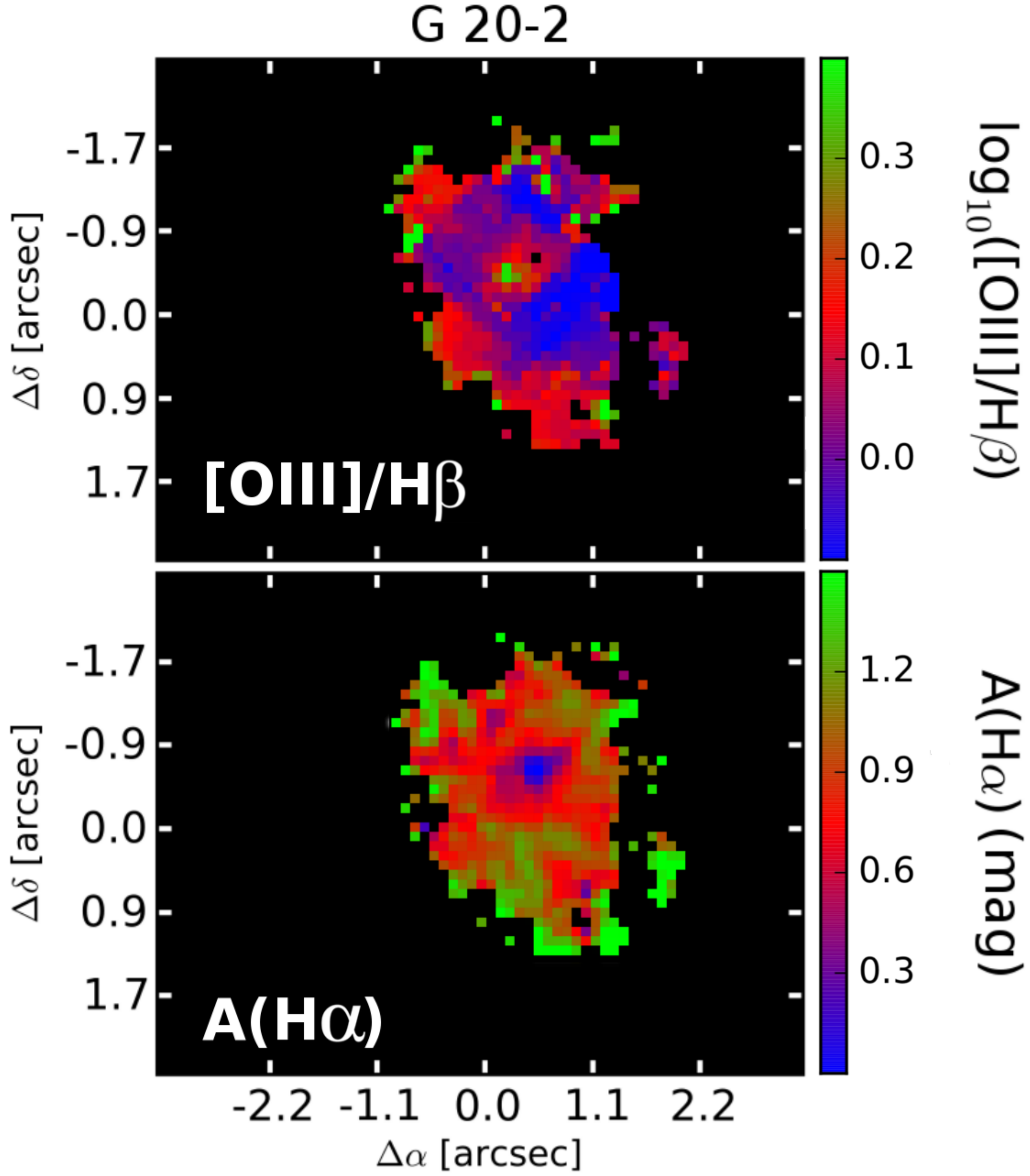}
  \caption{Comparison between the [OIII]/H$\beta$ ratio from GMOS observations and
    attenuation in galaxy G 20-2. This galaxy was found in integrated measurements to
    exhibit an excess of $E(B-V)_{stars}$ relative to $E(B-V)_{gas}$,
    which is generally unexpected. From the resolved observations of
    attenuation, the most striking feature is a drop in attenuation in
    the central region. Here we see that this drop corresponds to a
    point-like peak in [OIII]/H$\beta$ that may be indicative of a
    low-luminosity AGN.}\label{figure:ch4oratmaps}
\end{figure}

\vspace{2.5mm}
\noindent\textbf{D 13-5} is classified as a rotating disk from our initial SPIRAL
observations, and its continuum light follows an exponential profile, suggesting that this galaxy is an
undisturbed disk. H$\alpha$ photometry from HST reveals two irregularly
coiled and faint spiral arms extending away from a central ring of
strong star formation. The OSIRIS field of view is only slightly
larger than the nuclear starburst region, and thus, it is in this
region where we can reliably measure the Pa$\alpha$/H$\alpha$ ratio.

A by eye comparison between the maps of Pa$\alpha$ and H$\alpha$ flux
(Figure \ref{figure:ch4psfmm})
shows that the relative prominences of clumps in each map are
different. This is reflected in the relatively wide variation of 0.3 <
$A_{H\alpha}$ < 2.5
observed, with the largest values of $A_{H\alpha}$ $\simeq$ 2.5 found
for the most Pa$\alpha$ bright clump in the bottom right corner
(typical values in this area are 1.8 $\leq$ $A_{H\alpha}$ $\leq$
2.5). This contrasts with regions surrounding the most H$\alpha$
luminous clumps where typical values are 0.8 $\leq$ $A_{H\alpha}$
$\leq$ 1.5. The amount of variation in
$A_{H\alpha}$ observed in D 13-5 is the largest of any galaxy in this
sample, possibly due to its lower redshift ($z\sim0.07$) and thus higher spatial
resolution.

\vspace{2.5mm}
\noindent\textbf{G 04-1} was previously classified as a compact rotating disk,
and subsequent observations have revealed it to host an extremely
smooth exponential profile in its stellar light \citep{bassett14}. Line emission
is arranged in clumpy spiral arms as well as an extremely luminous
circumnuclear ring. The spiral arms observed in Pa$\alpha$ and
H$\alpha$ are mirrored by low surface brightness spiral arms
in the $K$-band continuum from our OSIRIS observations. 

We see little variation (standard deviation of $A_{H\alpha}$=0.35) between
the attenuation for clumps and the surrounding areas in the outer
disk. Relatively large
variations in the $A_{H\alpha}$ are seen in the central
star-forming ring. We find
an enhancement of the relative flux of Pa$\alpha$ in the top right
portion of the ring as pictured in Figure \ref{figure:ch4psfmm}
indicative of an increase in the column density of dust in this
region where the peak $A_{H\alpha}$ in a single spaxel is measured at
2.3 mags. Similarly, there is a relative decrease of Pa$\alpha$ on the
opposite side of the ring where we measure a minimum $A_{H\alpha}$ in
a single spaxel of 0.5 mag. The average difference in $A_{H\alpha}$ found in spaxels
between these two regions is $\sim$1 mag. This is consistent with one
portion of the ring being obscured by a relatively thick dust cloud while another
portion is shining through a less dusty region.  

\vspace{2.5mm}
\noindent\textbf{G 20-2} is an exponential disk with a number of
bright clumps situated in a star-forming ring surrounding a central
peak in H$\alpha$ flux \citep{bassett14}. We also find clumps at larger radii, the
brightest of which are near the detection limits of our OSIRIS IFS
observations. Overall we find a narrow range in the 
$A_{H\alpha}$, as seen in Figure \ref{figure:dusthists}, with nearly all spaxels within 0.5 magnitudes of
$A_{H\alpha}$=0.9 mag. The largest values are typically associated with
non-clump spaxels, however, this could be an effect of less reliable
Pa$\alpha$ fluxes in these regions. Compared to typical star-forming disks locally, this galaxy
can be considered to be quite average in its $A_{H\alpha}$ properties. 

Overall the levels of attenuation across the entire galaxy are similar
resulting in the clumpy
structure being completely absent from the map of $A_{H\alpha}$. The main
feature of the map of attenuation for G 20-2 is an apparent drop in
the amount of attenuation in the central region, including spaxels
with negative $A_{H\alpha}$ meaning that our measured line ratios are
\textit{smaller} than the intrinsic ratio suggested by case B
recombination (implying negative attenuation). From our GMOS-IFS
observations, we map [OIII]/H$\beta$, which is sensitive to
the ionization state of the gas \citep{baldwin81,kauff03c}, in Figure
\ref{figure:ch4oratmaps}. The extremely compact peak in [OIII]/H$\beta$ we
observe could be indicative of a low-luminosity AGN. 
Comparing the peak value of
$log_{10}([OIII]/$H$\beta)=0.38$ in the centre of G 20-2 with the
average values of $log_{10}([NII]/$H$\alpha)=-0.176$ from SPIRAL
observations places the centre of G 20-2 well within the AGN region of
the BPT diagram \citep{kauff03c}. 
We note that Pa$\alpha$ to H$\alpha$ ratios less than the
intrinsic ratio suggested by case B recombination have been measured
for AGN by \citet{kim10} consistent with the values found in the
central region of G 20-2.

\vspace{2.5mm}
\noindent\textbf{H 10-2} has been classified as a merging system based
on complex kinematics apparent in our GMOS-IFS observations. This is
consistent with the OSIRIS continuum morphology as H 10-2 is the only
galaxy in our sample that exhibits two peaks in continuum emission,
the cores of the two merging galaxies. The bulk of the line emission
is centered on these two continuum peaks, consistent with our merger
hypothesis as star-formation in mergers typically results from gas
funneling into the central regions \citep{scoville86,hopkinsp13}. Lower luminosity line emission is found
in a clumpy substructure associated with the brighter
(more massive) galaxy. This clumpy substructure is situated opposite
the less luminous continuum peak, which may indicate
that tidal forces have
triggered the clumpy star formation beyond the central region \citep[e.g.][]{toomre72,sanders86}.

From the histogram of spaxel $A_{H\alpha}$ measurements it is apparent
that this galaxy, consistent with the low metallicity from [NII]/H$\alpha$, has the lowest overall dust content. Similar to
galaxy G 20-2, we find the lowest values in the regions with the
largest continuum flux although variation from spaxel to spaxel in
this region can be quite large (0.0-0.66 mag). Clumps show less
variation with values falling near 0.5 mag, slightly higher than the
global average. A scenario for the formation of this galaxy is an ongoing, gas-poor \citep[and therefore low dust,][]{cortese12b}
merger. Although the gas fraction is relatively low \citep{fisher14},
the available gas can be funneled inward fueling a large
SFR. \citet{brandl05}
present IR observations of the Antennae galaxies, a famous merger,
finding variation in attenuation from 0-9 magnitudes. Their false
colour images reveal large, and very blue, star-forming regions
consistent with intense and relatively unobscured star-formation as
well as thick dust clouds. This is roughly consistent with our
observations of H 10-2 assuming we are insensitive to low levels of
highly obscured star-formation that may be present. 

\subsubsection{Clump-to-Clump Measurements of $A_{H\alpha}$}\label{section:ch4clmeas}

\begin{figure}{
    \centering
    \includegraphics[width=\columnwidth]{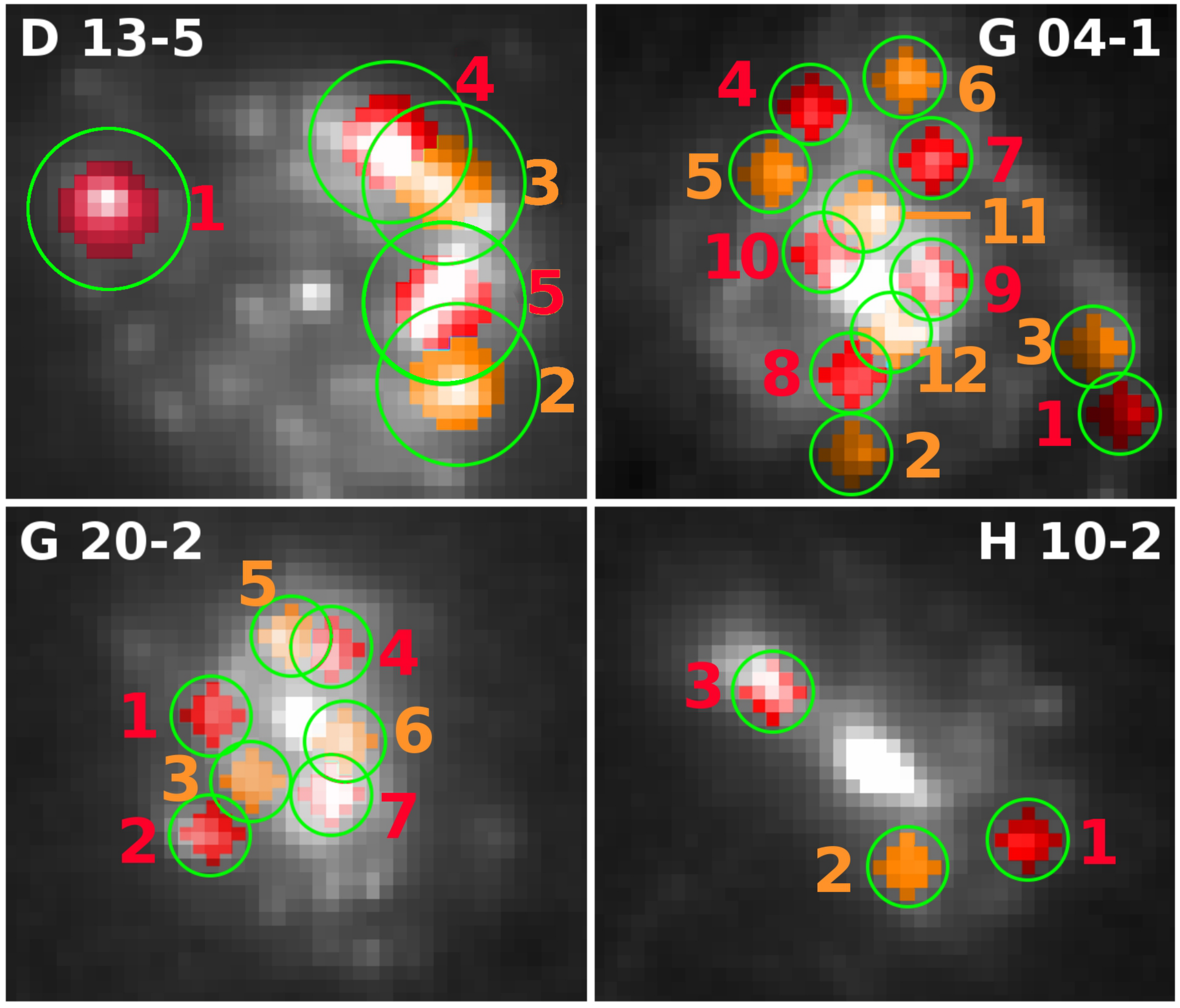}
    \caption{Apertures used to define our clump $F($Pa$\alpha)$ and
      $F($H$\alpha)$. In each panel we show the H$\alpha$ image of the
      indicated galaxy with green circles around each clump studied
      here. We alter the colouring of pixels in the images to red or orange
      to indicate exactly which pixels are selected. Red and orange
      colours are arbitrary and alternate in order to clearly
      distinguish between pixels included in clumps that are nearby
      one another. Clumps are numbered in order of increasing
      H$\alpha$ flux in order to easily identify them in Figures
      \ref{figure:ch4clfs} and
      \ref{figure:ch4clavs}}\label{figure:ch4clumplocs}
    }
\end{figure}

\begin{figure}
  \includegraphics[width=\columnwidth]{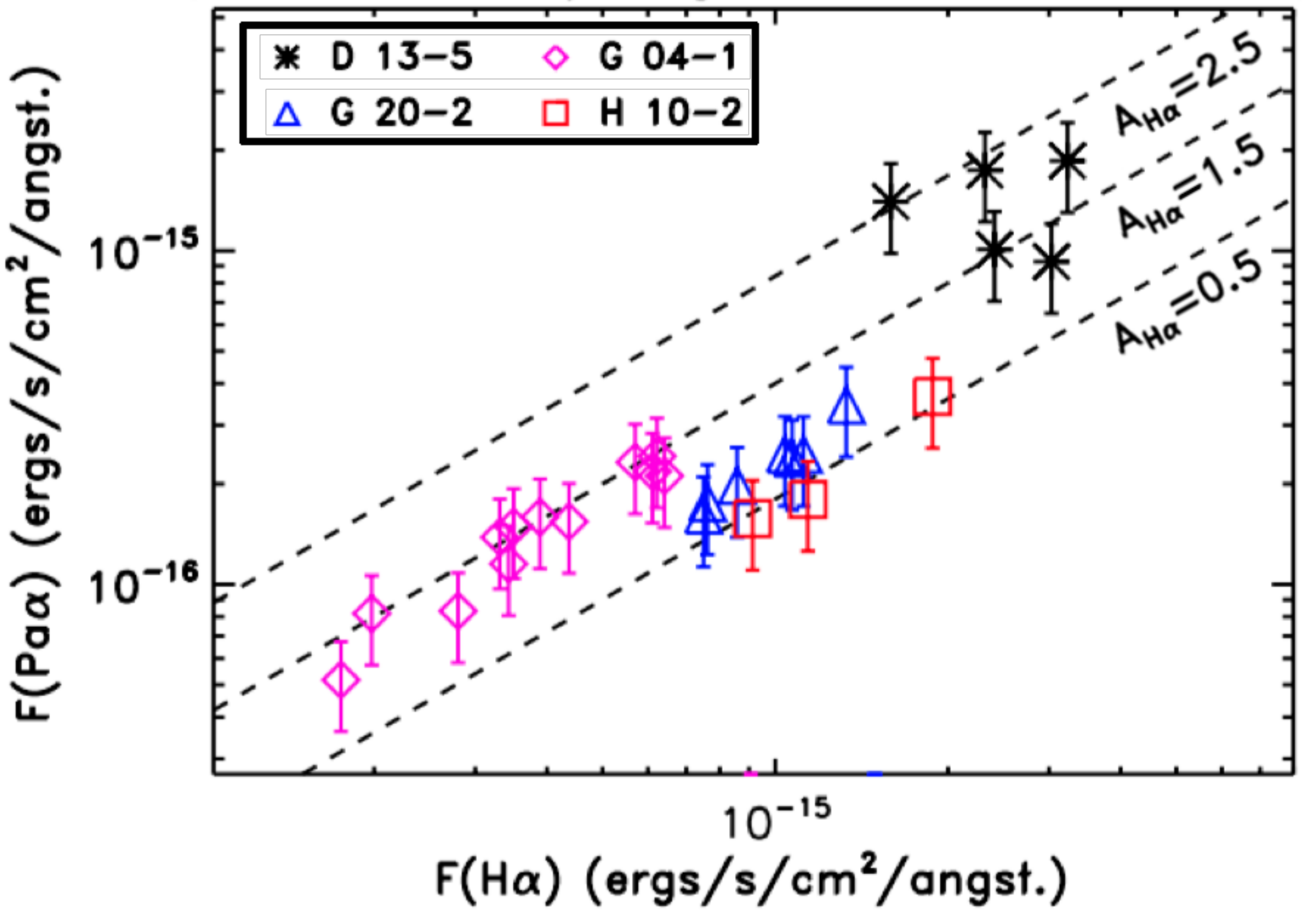}
  \caption{The relation between Pa$\alpha$ and H$\alpha$ measured in
    apertures centered on each clump. Results for each galaxy are
    plotted using different colours and symbols. Flux ratios
    corresponding to various constant levels of attenuation are
    indicated with dashed lines. Three galaxies appear to follow these
  trends while D 13-5 displays a larger scatter. Error bars reflect
  the $\sim$30\% error in our OSIRIS flux calibration, most
likely a systematic effect.}\label{figure:ch4clfs}
\end{figure}

In this section, we compare H$\alpha$ versus Pa$\alpha$ flux
for clumps in each of our galaxies. Note that in this analysis we omit
central clumps as the properties of these clumps will be influenced by
different physics than other clumps such as strong gas inflows
associated with their being coincident with the minima of the galactic
gravitational potential. For galaxies in the redshift range
0.129 < $z$ < 0.151 we measure fluxes in apertures of
0$\farcs$3, while D 13-5, which is at $z=0.07535$, we use 
apertures of 0$\farcs$6 to account for the roughly $2\times$ higher spatial resolution we
achieve. Our apertures are illustrated in Figure
\ref{figure:ch4clumplocs} with clumps numbered in order of increasing
H$\alpha$ flux. This analysis represents a compromise between resolving clumps
and minimising the PSF effects. These apertures correspond to physical sizes of 0.86, 0.75,
0.70, and 0.79 kpc for galaxies D 13-5, G 04-1, G 20-2, and H 10-2
respectively. 

We do not attempt to background subtract
the flux in each aperture, which, in theory, could account for
contamination from diffuse line emission. Due to the inhomogenous,
clumpy nature of the emission line flux in these galaxies, a
background subtraction such as this would be difficult to estimate and
would likely only add to the uncertainty of our measurements. 

Flux ratios measured in apertures in this way provide
more robust $A_{H\alpha}$ estimates when compared to a spaxel by
spaxel analysis, allowing us to clearly show if there is significant
variation in attenuation from clump to clump. We first plot H$\alpha$
versus Pa$\alpha$ flux for clumps in
Figure \ref{figure:ch4clfs}.
For galaxies G 04-1, G 20-2 and H 10-2 we
  find a correlation between F(H$\alpha$) and F(Pa$\alpha$) 
  suggesting that
  the attenuation suffered by different clumps in a given galaxy is quite
  similar (F(Pa$\alpha$)/F(H$\alpha$) = constant). We indicate with
  dashed lines the relation corresponding to fixed $A_{H\alpha}$ of
  0.5, 1.5 and 2.5 mags.

Galaxy D 13-5 exhibits an apparent lack of correlation between
F(H$\alpha$) and F(Pa$\alpha$). We believe that this not an effect of
the increased spatial resolution achieved for this galaxy as we have
accounted for this using a larger aperture for D 13-5. Rather, we
suggest that this is an effect of larger scale variation in dust
geometry, such as dust lanes, which appear in Figure
\ref{figure:dustmaps} to obscure clumps 2 and 5 significantly more
than clumps 3 and 4. We also see in Figure
\ref{figure:ch4clfs} that clumps 2 and 5 have similar attenuation, and
the same is true of clumps 3 and 4, consistent with this picture. In
the other three galaxies, we observe clumps that are separated by
distances much larger than our PSF scale, meaning that large
clump-to-clump variations like those observed in D 13-5 should be
apparent in Figure \ref{figure:dustmaps} if present. As we noted in
the previous subsection, the $A_{H\alpha}$ map of the centre of G 04-1
exhibits similar variation as that of D 13-5. Unlike D 13-5, however, the locations of
the minimum and maximum are not coincident with the locations of
clumps, which explains why this variation is not apparent in Figure
\ref{figure:ch4clfs}. 

\begin{figure}
  \centering
  \includegraphics[width=\columnwidth]{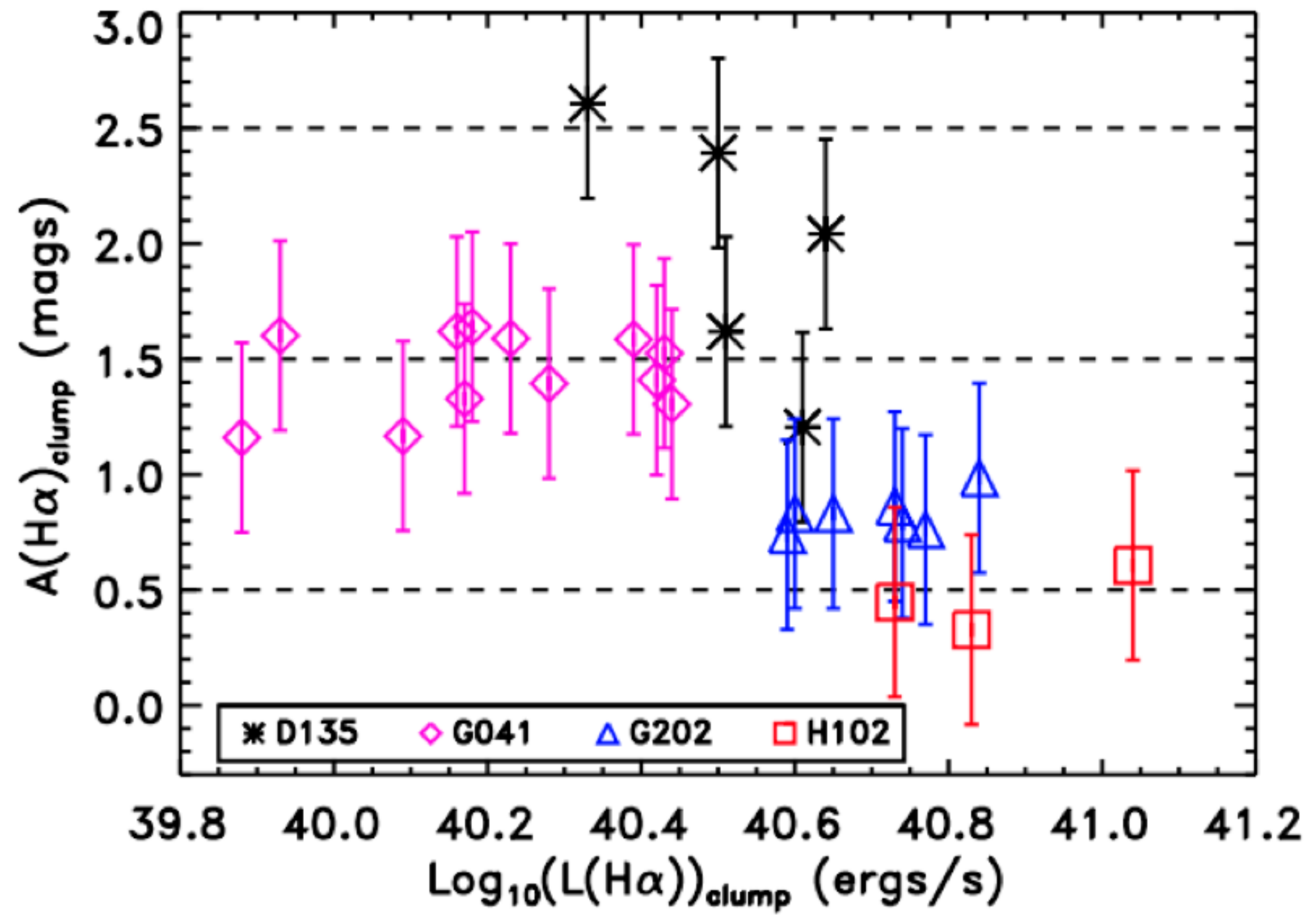}
  \caption{For each clump we plot the H$\alpha$ aperture luminosity
    versus the attenuation suffered at the wavelength of H$\alpha$
    with symbols corresponding to each galaxy. In galaxies G 04-1, G
    20-2 and H 10-2 we see
  little internal variation in the value of $A_{H\alpha}$, with
  larger differences seen when comparing average values between galaxies. In
  comparison, D 13-5 shows
  larger differences in the amount of attenuation even at fixed
  L(H$\alpha$), suggesting a clumpier distribution of dust. Errors on
  $A_{H\alpha}$ reflect the $\sim$30\% error on
  our OSIRIS flux calibration, most likely a systematic effect.}\label{figure:ch4clavs}
\end{figure}

The apparent lack of variation of $A_{H\alpha}$ measured in clumps for a given
galaxy can also be seen in Figure \ref{figure:ch4clavs} where we plot the
luminosity of H$\alpha$ versus $A_{H\alpha}$ calculated
as described in Section \ref{section:ch4bdcorr}. Figure
\ref{figure:ch4clfs} shows that clump H$\alpha$ luminosity and
attenuation are not correlated in galaxies G 04-1, G20-2, and H 10-2.
We also observer a narrow spread in clump $A_{H\alpha}$ with values comparable
to observations of local star-forming galaxies
\citep{keel01,matthews01,takeuchi05,cortese08}. Galaxy D 13-5 exhibits
the widest range in clump-to-clump $A_{H\alpha}$ as shown in
Figure \ref{figure:ch4clavs}. The two clumps with lower $A_{H\alpha}$
are numbers 3 and 4 in Figure \ref{figure:ch4clumplocs}, consistent with
a visual inspection of H$\alpha$ and Pa$\alpha$ maps presented in
Figure \ref{figure:ch4psfmm}. G 04-1 has relatively uniform
clump-to-clump $A_{H\alpha}$ with the exception of clumps 1 and 3,
which are less attenuated than the others. From Figure
\ref{figure:ch4clumplocs} we see that these fall towards the end of the
most prominent spiral arm, suggestive a low optical depth of dust at
large radii. The remaining galaxies, G 20-2 and H 10-2, exhibit the
smallest variation from clump to clump suggesting similar dust content
across the observed regions.

\section{Discussion}\label{section:ch4discussion}

\subsection{Integrated Versus Resolved Attenuation}

In Section \ref{section:intav} we investigated the integrated
properties of the full DYNAMO sample presented by \citet{green14}. We
found that highly star-forming DYNAMO galaxies are well matched to the
high redshift MOSDEF sample of \citet{reddy15} when considering the $M_{*}$
vs $E(B-V)_{gas}$ and SFR vs $E(B-V)_{gas}$ relationships. We also
compare $E(B-V)_{gas}$ vs $E(B-V)_{stars}$ for these two samples
finding many  galaxies to be in agreement with the relationship
$E(B-V)_{stars} = 0.44 \times E(B-V)_{gas}$ found for local starbursts
\citep{calz97}. At higher
attenuation, however there does appear to be a trend for both DYNAMO
and MOSDEF galaxies to fall below this line, closer to the relation
$E(B-V)_{stars} = E(B-V)_{gas}$. Fitting a linear relationship to
DYNAMO and MOSDEF data we find the relationships $E(B-V)_{stars} =
0.78\pm0.08 \times E(B-V)_{gas}$ and $E(B-V)_{stars} = 1.07\pm0.25
\times E(B-V)_{gas}$ respectively. Such behavior also been seen in other
high redshift samples \citep{erb06c,reddy10,kashino13}. This difference
may relate to sSFR as shown by \citet{wild11} and \citet{price14},
however these results are based on stacking of observations of many
galaxies. \citet{reddy15}, who select galaxies from the same parent
sample as \citet{price14}, show that there is a significant scatter in
$E(B-V)_{gas}$ vs $E(B-V)_{stars}$, independent of many other galaxy
properties, which may limit the usefulness of such stacking
analyses. 

In general, the fact that emission from ionized gas is $\sim$2 times
as attenuated as the light from stars can be explained by the fact that
stars migrate away from dusty star-forming regions over time while
ionized gas is, by necessity, always associated with regions of
intense star-formation \citep{charlot00,calz01,wild11}. Simple, screen-like geometries
between stars, gas, and dust are typically assumed where stars and gas
are both attenuated by a diffuse dust component while the gas is
attenuated by an additional dust screen associated with star-forming
regions. For galaxies G 20-2 and H 10-2 we find a relatively smooth
$A_{H\alpha}$ distributions suggesting
that such a simple picture may be appropriate. Galaxies G 04-1 and D
13-5, exhibits a slightly larger dispersion in $A_{H\alpha}$, which
indicates that large scale variations in dust column density, i.e. a
clumpy distribution of dust. The maxima and minima in attenuation for
galaxy D 13-5 are spatially correlated with locations of clumps
leading to a relatively large clump-to-clump variation in
$A_{H\alpha}$. In contrast, maximum and minimum attenuations in G 04-1
do not correspond with clump locations, leading to little variation in
clump-to-clump $A_{H\alpha}$, similar to G 20-2 and H 10-2 (see Figure
\ref{figure:ch4clavs}). 

Using resolved maps of attenuation we can ask: what is the difference in integrated SFR when a single $A_{H\alpha}$
is assumed versus a spatially resolved $A_{H\alpha}$ correction? To do
this we measure the total attenuation corrected H$\alpha$ flux from our HST observations in
two ways: first using a single integrated attenuation correction for
H$\alpha$, taken from
the MPA-JHU VAC values, and second by correcting individual pixels based on our $A_{H\alpha}$ maps.
For galaxies D 13-5, G
04-1, G 20-2, and H 10-2 we measure fractional differences,
$F($H$\alpha_{int})-F($H$\alpha_{map})$/$F($H$\alpha_{int})$, of +0.28,
-0.05, +0.03, and +0.12 respectively. Thus we find using an integrated
attenuation 
corrections that the SFR in these four galaxies ranges from a 5\%
under prediction to a 28\% over prediction of the SFR compared to
resolved attenuation corrections, with no systematic trend between
galaxies. 

The largest difference between these two SFR estimates is found for our
lower redshift galaxy, D 13-5. This galaxy also exhibits a large
variation in $A_{H\alpha}$ from clump to clump, which we attribute to
large scale dust
features. Clumps in galaxies G 20-2 and H 10-2 have large enough
separations that we should resolve similar differences in attenuation
if they are present. In G 04-1 we observe a similar variation in our
$A_{H\alpha}$ map, however maximum and minimum attenuations do not
correlate with clumps meaning assuming a flat $A_{H\alpha}$ correction
will have less of an effect on the integrated SFR. This may also account for the fact that in
integrated measurements galaxies
G 04-1, G 20-2, and H 10-2 fall on the $E(B-V)_{stars}=0.44 \times
E(B-V)_{gas}$ relation while D 13-5 sits significantly below this, as an
assumption of a simple screen dust geometry is implicit
in these calculations. This assumption may not be appropriate for a
clumpy distribution of dust that attenuates different star-forming regions at different
levels, which may be more representative of D 13-5. Of course, from a sample of
four galaxies it is not possible to extrapolate to the full sample of
clumpy high redshift galaxies, again highlighting the necessity of
larger samples of resolved attenuation observations. The
observation that SFRs in DYNAMO galaxies may be biased and
overpredicted could have important implications regarding the
normalisation of the main sequence of star-forming galaxies at high
redshift. 

\subsection{An Absence of Highly Attenuated Clumps}

The main advantage of using IR emission lines for studying attenuation
in galaxies is the low intrinsic attenuation at these
wavelengths allowing lines such as Pa$\alpha$ to emerge from deep
within dust-enshrouded regions. This means that using observations of Pa$\alpha$ will be
significantly more sensitive to regions of high dust obscuration than
optical and UV techniques such as the Balmer decrement or the UV slope,
$\beta$ \citep{meurer99}. For foreground dust geometries the implication is that
optical techniques will be biased towards unextincted regions and may
not be sensitive to star formation located behind highly
obscuring dust clouds \citep{calz97}.

One key result from our HST+OSIRIS sample, which is apparent from Figure
\ref{figure:ch4psfmm}, is a lack of extremely bright clumps in our
Pa$\alpha$ maps that are absent (or at least very low luminosity) in
our H$\alpha$ imaging. This would be indicative of star
formation that is sitting behind a very high column density region of
dust, which completely attenuates the optical emission but allows the
IR emission to shine through.
\citet{genzel13} have performed semi-resolved
observations of CO (3-2) in a single main-sequence galaxy at $z=1.53$
from the PHIBSS survey \citep{tacc10,tacc13}, finding evidence of
large quantities of molecular gas spatially correlated with
their map of $A_{V}$ (from SED fitting). From this observation they
estimate that, assuming the dust is situated in a single foreground
cloud, $A_{V}$ in these regions could reach as high as
$\sim$50 mag. Assuming a \citet{card89} dust curve this implies an
attenuation of $\sim8$ mag even at the wavelength of Pa$\alpha$. We estimate that an individual
star-forming clump situated behind such a cloud would require a SFR
$\sim$50 $M_{\odot}$ yr$^{-1}$ to be detected by our OSIRIS
observations of our most nearby galaxy, thus we can not rule out the
possibility of lower levels of highly obscured star-formation in
DYNAMO galaxies. In local star-forming galaxies, however, dust is most
likely clumped on sub-kpc scales and/or mixed with stars and gas
\citep[e.g.][]{bedregal09,liu13,piqueras13,boquien15}. Thus highly
star-forming \citep[and therefore large,][]{wisnioski12} clumps are unlikely to be fully obscured
by high column density dust clouds. Furthermore, we observe a maximum H$\alpha$
attenuation of $\sim$3 mag thus a jump from this to regions of
$A_{H\alpha}$ > 10 with no intermediate cases is unlikely. For these reasons we suggest that the levels of highly attenuated
star-formation in clumpy DYNAMO galaxies are minor relative to that
currently observable in clumps.

These result may carry implications regarding observations of
clumpy main-sequence galaxies at high redshift \citep{wright09,forster09,wisnioski11,epinat12,swinbank12,wisnioski14}. Resolving clumps in
$z\sim1-2$ star-forming galaxies in the IR, as is done here for DYNAMO galaxies
at $z\sim0.1$, is not possible using
current facilities. If the physical
conditions of DYNAMO galaxies can be considered as similar to the
conditions in clumpy high-$z$ disks, then it should be reasonable to
assume that observations of H$\alpha$ can, in some cases, provide a full
census of star-forming regions in a given galaxy. We stress again,
though, that this result
is based on observations of four galaxies, thus, it is not reasonable to
extrapolate to large samples of star-forming galaxies at
high-redshift. Future observations probing the IR regime of high
redshift galaxies such as those performed using the James Webb Space
Telescope will further shed light on these issues.

\subsection{Spatial Variation in Attenuation}\label{section:ch4spva}

\begin{table}[t]
  \begin{center}
  \caption{Average Clump and Non-Clump Attenuation Values}
  \vspace{2mm}
  \begin{tabular}{ c c c c }
    \hline\hline
    ID & <$A_{H\alpha}($clump$)$>\footnote{mean value of $A_{H\alpha}$
         in spaxels within all clump apertures} & <$A_{H\alpha}($non-clump$)$>\footnote{mean value of $A_{H\alpha}$
         in spaxels outside of clump apertures} & $A_{H\alpha,
                                                  int}$\footnote{integrated
                                                  $A_{H\alpha}$ based
                                                  on SDSS measurements
                                                  of the Balmer decrement}\\
    &  mag & mag & mag\\
    \hline
    D 13-5 & 1.80$\pm$0.52 & 1.65$\pm$0.49 & 1.46\\
    G 04-1 & 1.52$\pm$0.26 & 1.41$\pm$0.37 & 1.55\\
    G 20-2 & 0.89$\pm$0.14 & 0.98$\pm$0.32 & 0.86\\
    H 10-2 & 0.35$\pm$0.24 & 0.44$\pm$0.44 & 0.21\\ 
    \hline
  \end{tabular}\label{table:avav}\\
  \end{center}
\end{table}

We observe a range of attenuation values both between galaxies as well
as within individual galaxies with typical values 0.0 < A$_{H\alpha}$
< 3.0 (see Figure \ref{figure:dusthists}). This result is generally consistent with attenuation
measurements of star bursting galaxies at both low \citep{calz00} and
high redshifts \citep{forster11}. 
This result is also consistent with the IFS observations of
\citet{boquien09} who find A$_{V}$ values of $0.1-1$ for a sample of low SFR, face-on disks. Highly attenuated LIRGS and
ULIRGS on the other hand, are found to have typical median A$_{V}$ values of 4-6
mag with individual measurements ranging from $\sim$1 to $\sim$20 mags within a
single galaxy \citep{bedregal09,piqueras13}. The general result is that
dust is geometrically clumpy, and galaxies with larger overall
dust content host the highest column density clumps. Confirming this
observation however will rely upon measurements of
attenuation in larger samples of galaxies of various types with sub-kpc resolution. 

The fact that we observe a large spread in values for galaxy D
13-5 may be related to the lower redshift of this object when
compared with the remainder of the sample. This redshift distance
corresponds to an increased spatial resolution for this object of
863 pc when compared to $\sim$1.5 pc at high redshift. This is
consistent with the study of \citet{boquien15} who test the effects
of resolution on the resolved attenuation measurements of M33. The
general result of this study is that spatial variation in attenuation
maps becomes more apparent for small spatial sampling scales, with
differences reducing to $\sim0$ at scales of $\sim$1 kpc. We perform a
simple test by rebinning the $A_{H\alpha}$ map of D 13-5 by a factor
of two and find the same spread of values as in the unbinned
image. The significant enhancement in $A_{H\alpha}$ at the bottom
right of D 13-5 relative to the rest of the galaxy is still apparent,
suggesting differences in attenuation seen in Figure
\ref{figure:dustmaps} is due to large scale variation in dust covering
fraction (e.g. a dust lane). We have also pointed out that the OSIRIS
observations of D 13-5 cover only the central star-forming ring that
can be considered comparable to the central region of G 04-1 observed at a
higher spatial resolution. In the
star-forming ring of G 04-1 we observe a similarly large variation in
attenuation giving further evidence that the observed variation in
attenuation is not an effect of resolution. In G 04-1, this variation does not correlate with
clumps, thus clump-to-clump measurements, and likely integrated
measurements, of SFR are less affected by variable attenuation in G
04-1 when compared with D 13-5. 

The spatial resolutions probed in this study are roughly
comparable to the measured size of clumps in these galaxies. This
means that while resolving dust substructure within individual clumps
is not possible, differences in attenuation from clump to clump should
be apparent if present. It is therefor remarkable that in each galaxy
there is such a small variation in attenuation from clump to clump as
shown by Figures \ref{figure:ch4clfs} and \ref{figure:ch4clavs}. 
In particular, galaxies G 04-1, G 20-2, and H 10-2 each exhibit a very
flat relationship between clump H$\alpha$ luminsoity and
$A_{H\alpha}$. This may be due to the turbulent nature of these
galaxies that can result in
significant mixing of stars, gas, and dust. A similar result is found
by \citet{kreckel13} for galaxy NGC 2146, the strongest starburst
system in their sample. We note, however, that
comparing clump and non-clump spaxels in our DYNAMO galaxies using a two sample KS-test
rejects the hypothesis that $A_{H\alpha}$ values are taken from the
same parent distribution. This is most likely due to the fact the
largest $A_{H\alpha}$ values are found in 
  non-clump spaxels, resulting from less reliable 
  Pa$\alpha$ flux measurements outside of bright clumps. Figure
  \ref{figure:dusthists} shows that the
  average values of $A_{H\alpha}$ between clump and non-clump regions
  are very similar.

A flat distribution of $A_{H\alpha}$ in turbulent DYNAMO galaxies is
consistent with models of the evolution of clumpy star formation
such as those of \citet{bour14} who argue for long-lived
clumps that experience both strong inflows and outflows. Such
activity will greatly facilitate the mixing of material between clumps
and between clumpy and non-clumpy regions. 
Such a model naturally
results in relatively flat spatial distributions in $A_{H\alpha}$. We
do find significant variation in the average attenuation of clumps
from galaxy to galaxy, however this is likely due to differences in
stellar mass. This is consistent with our investigation of integrated
attenuations as we find the four galaxies in our resolved attenuations
sample to follow a trend of increasing $E(B-V)_{gas}$ with increasing
mass in Figure \ref{figure:ch4agascorr}. 

\section{Conclusions}\label{section:ch4summary}

In this paper, we have presented integrated attenuation measurements
of a sample of 67 DYNAMO galaxies from SDSS observations as well as a
resolved study of attenuation in four highly star-forming
DYNAMO galaxies.
The latter is achieved by combining high-resolution H$\alpha$
imaging from HST with AO-assisted IFS from OSIRIS at Keck targeting
the Pa$\alpha$ emission line in the IR. By utilising emission at long
wavelengths we are sensitive to highly attenuated star-formation which
may be missed by observations of optical emission lines. The
conclusions of our analyses are as follows:

\begin{itemize}

  \item From integrated observations we find that DYNAMO galaxies
    exhibit a larger spread in integrated SFR at fixed
    $E(B-V)_{gas}$ than the high redshift MOSDEF sample
    \citep{reddy15}, however, considering only highly star forming
    DYNAMO galaxies alone there is much better agreement.

  \item Comparing integrated $E(B-V)_{gas}$ and $E(B-V)_{stars}$ for
    DYNAMO and MOSDEF we find a similar trend that, at high
    attenuation, galaxies deviate from the local starburst relation
    $E(B-V)_{stars}=0.44 \times E(B-V)_{gas}$ moving towards
    $E(B-V)_{stars}=E(B-V)_{gas}$. We fit linear relationships to
    DYNAMO and MOSDEF galaxies finding $E(B-V)_{stars}=0.78\pm0.08 \times
    E(B-V)_{gas}$ and $E(B-V)_{stars}=1.07\pm0.25 \times E(B-V)_{gas}$ respectively. For DYNAMO galaxies this does not
    seem to depend on SFR, however. 

  \item In four DYNAMO galaxies we find no evidence for highly attenuated and strongly
    star-forming clumps, which would be readily apparent in our Pa$\alpha$
    observations. This does not preclude the possibility of lower
    levels of highly obscured star-formation, however, such
    star-formation would likely be negligible when compared to the
    extreme SFRs of observed clumps. 

  \item We find mild spatial variation in the amount of
    attenuation depending on the spatial location within a given
    galaxy. Values of $A_{V}$ vary from $\sim0$ to $\sim3$ with most
    values in the 0.5-1.5 range, typical of local star-forming
    galaxies. Within a single galaxy, the spread in values is $\sim$1
    mag.

\end{itemize}

The latter two conclusions, from our resolved observations, suggest
that the bulk of the star-formation present
in these galaxies is already accounted for in our H$\alpha$ imaging
even though it will be more affected by the presence of dust. We
stress, however, that this result is based on observations of four
galaxies and they may not be immediately applicable to the general population
of clumpy, star-forming galaxies at $z>1.5$. In future work we plan on
investigating the integrated attenuations of DYNAMO galaxies by
comparing $E(B-V)_{gas}$, from the Balmer decrement, and
$E(B-V)_{stars}$, computed through full SED fitting including far-IR
data constraining dust emission. Work investigating the direct
detection of emission from dust in DYNAMO galaxies is currently underway using
observations from the Wide-field IR Survey Explorer
\citep[WISE,][]{wright10} that will compare dust masses with molecular
gas masses \citep[from CO detections using PdBI,][White et al. in
prep]{fisher14}. These analyses will
help us to test the applicability of local calibrations
\citep[e.g.][]{calz97} to star-forming galaxies at high redshift and
how the relative attenuations suffered by stars and gas relate to
other galaxy properties. 

\vspace{5mm}

The authors would like to thank Naveen Reddy for providing data from
the MOSDEF project for comparison to our observations as well as
Katinka Ger\'{e}b for constructive comments in the preparation of this
manuscript. We would also like to thank the anonymous referee for
helping to improve the readability of this work.
Support for this project is provided in part by the Victorian
Department of State Development, Business and Innovation through the
Victorian International Research Scholarship (VIRS). This research was
supported under Australian Research Council's Discovery Projects
funding scheme (project number DP130101460). 
(Some of) The data presented herein were obtained at the W.M. Keck
Observatory, which is operated as a scientific partnership among the
California Institute of Technology, the University of California and
the National Aeronautics and Space Administration. The Observatory was
made possible by the generous financial support of the W.M. Keck
Foundation. The authors wish to recognize and acknowledge the very
significant cultural role and reverence that the summit of Mauna Kea
has always had within the indigenous Hawaiian community.  We are most
fortunate to have the opportunity to conduct observations from this
mountain. Portions of this
work are based on observations obtained at Gemini Observatory
(GS-2011B-Q-88, processed using the Gemini IRAF package), which is
operated by the Association of Universities
for Research in Astronomy, Inc., under a cooperative agreement with
the NSF on behalf of the Gemini partnership: the National Science
Foundation (United States), the National Research Council (Canada),
CONICYT (Chile), Ministerio de Ciencia, Tecnolog\'{i}a e
Innovaci\'{o}n Productiva (Argentina), and Minist\'{e}rio da
Ci\^{e}ncia, Tecnologia e Inova\c{c}\~{a}o (Brazil).
This research has made use of data from HRS project. HRS is a Herschel
Key Programme utilising Guaranteed Time from the SPIRE instrument
team, ESAC scientists and a mission scientist.  
The HRS data was accessed through the Herschel Database in Marseille
(HeDaM - http://hedam.lam.fr) operated by CeSAM and hosted by the
Laboratoire d'Astrophysique de Marseille. 




\bibliographystyle{mnras}
\bibliography{refs} 



\appendix

\section{Testing the Effects of PSF Matching}\label{section:ch4psfmtest}

\begin{figure}
\includegraphics[width=\columnwidth]{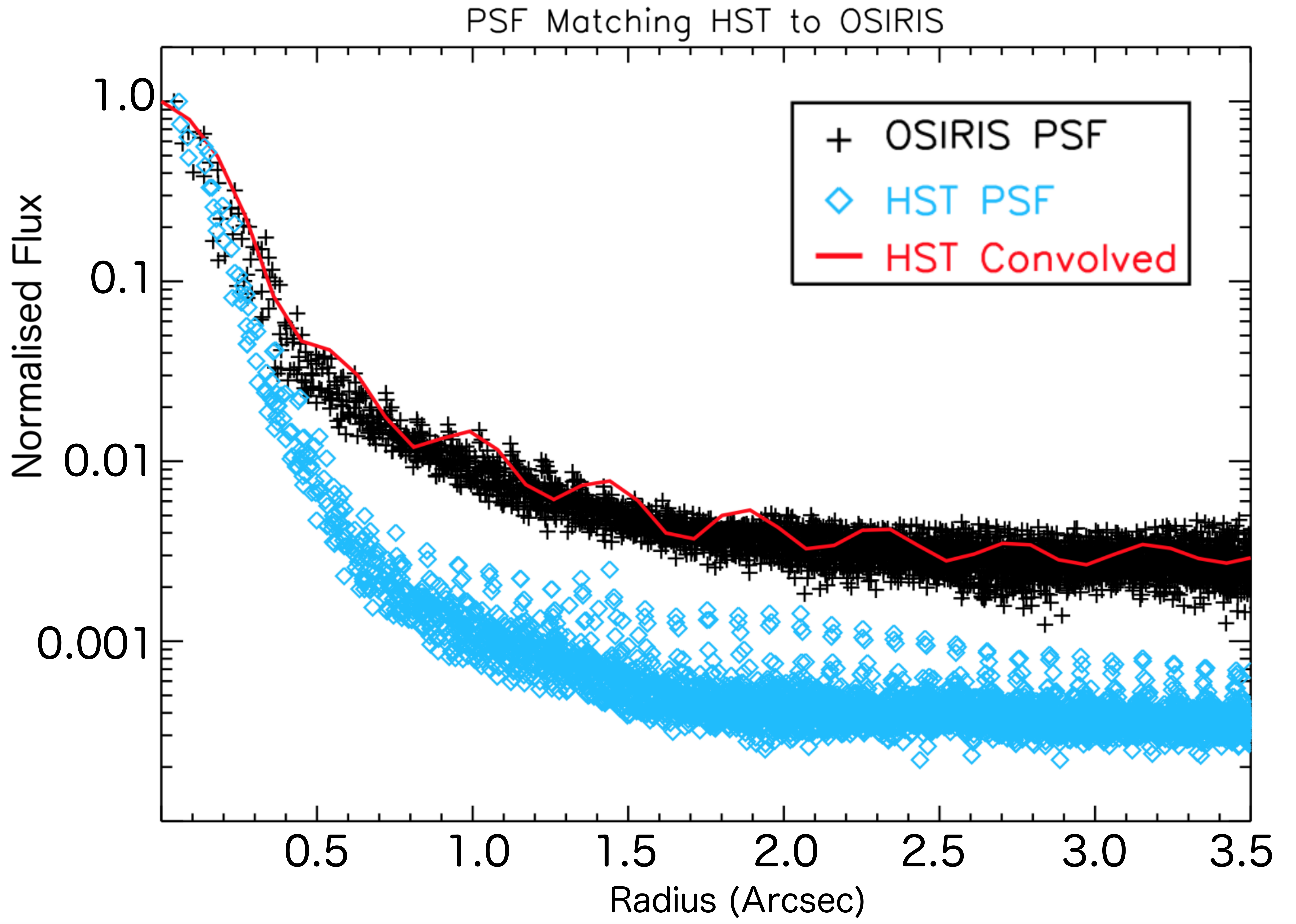}
\vspace{0pt}
 \caption{The results of our PSF matching procedure. We plot the input
   data for our empirical OSIRIS (black crosses) and HST (blue
   diamonds) PSFs normalized to their respective central fluxes. Each
   PSF was constructed from observed stars during our program as
   described in Section \ref{section:ch4psfmatch}. Overplotted in red is
   the binned radial profile of our output HST PSF after being matched
   to OSIRIS using the IRAF task \textit{psfmatch}. The oscillations
   apparent in the convolved profile are a result of Fourier
   Transforms used by \textit{psfmatch} in combination with the small
   field of view of OSIRIS.} \label{figure:ch4psfmatch}
\end{figure}

We would like to understand any possible biases that may be
introduced into our observed maps of $A_{H\alpha}$ as a result of our PSF
matching procedure. We illustrate the difference between the HST and
OSIRIS PSFs, as well as the result of applying our PSF matching
procedure directly to our HST PSF, in Figure
\ref{figure:ch4psfmatch}. The PSF effects here may be quite complex
due to the bright PSF wings and rapid temporal variation. Furthermore,
these effects will
depend on the relative proximity of bright clumps as well as the
relative brightness of clump versus disk light. To test these effects we create
two additional maps of $A_{H\alpha}$ 
for each galaxies and compare these to the maps presented in Section
\ref{section:ch4dustmaps}. Together, these make up our three test
cases described below:

\vspace{2.5mm}

\noindent \textbf{Fiducial Map:} As a starting point for this test we
take the maps presented in Section \ref{section:ch4dustmaps}. To be
explicit here, these maps are produced by performing the IRAF task
\textit{psfmatch} using empirical PSFs for our HST and OSIRIS
observations and registering the images as described in the text. We divide our OSIRIS Pa$\alpha$
map by our PSF matched HST H$\alpha$ map and convert this to
$A_{H\alpha}$ using Equation \ref{equation:paha}. As this is the
method used in our analysis, we will refer to this as the fiducial
case.

\vspace{2.5mm}

\noindent \textbf{PSF Only:} In this instance we would like to isolate
the effects of our PSF matching procedure. This is done by producing
an artificial Pa$\alpha$ map from our H$\alpha$ observations. To
achieve this we perform the PSF matching procedure described in the
text on our HST H$\alpha$ image, then scale it by a factor determined
from Equation \ref{equation:paha} to give a value of $A_{H\alpha}$ of 1 mag. We
then divide this artificial Pa$\alpha$ map by the original (non-PSF
matched) HST H$\alpha$ map and convert to $A_{H\alpha}$ using Equation
\ref{equation:paha}. Had we not performed a PSF matching step, this
would return a flat $A_{H\alpha}$ map with a value of 1 (by design),
thus any variation in the produced map will be \textit{entirely due to
  our PSF matching procedure.}

\vspace{2.5mm}

\noindent \textbf{Unmatched:} Finally we create a map of
$A_{H\alpha}$ applying no PSF matching to either map
whatsoever. We simply register our unmatched HST H$\alpha$ map to our
OSIRIS Pa$\alpha$ map without PSF matching, divide this by our
Pa$\alpha$ map, and convert to $A_{H\alpha}$ using Equation
\ref{equation:paha}. By comparing this case with the fiducial map
described above we can isolate the effect our PSF matching has on the
maps presented in Section \ref{section:ch4dustmaps}. 

\vspace{2.5mm}

\begin{figure}
  \includegraphics[width=\columnwidth]{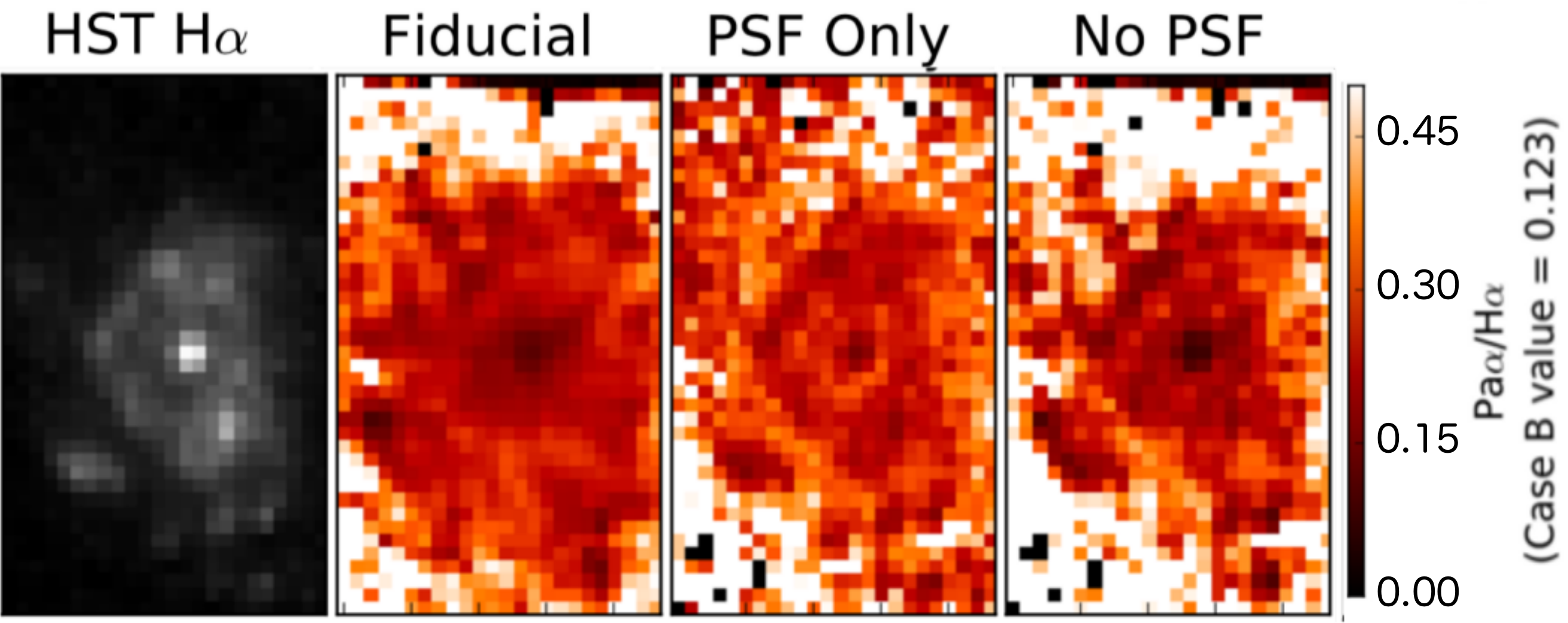}
  \caption{Maps of Pa$\alpha$/H$\alpha$ for galaxy G 20-2 showing the
    three test cases described in Appendix
    \ref{section:ch4psfmtest} as well as the HST map of H$\alpha$
    (without PSF matching). From left to right this shows the
    H$\alpha$ map, our fiducial case, PSF matching only, and no PSF
    matching. The intrinsic value of this ratio assuming Case B
    recombination is 0.123, thus larger values indicate regions where
    $A_{H\alpha}$ > 0. The two test cases show a correspondence between
    H$\alpha$ peaks (clumps) and minima in $A_{H\alpha}$, while clumpy
  substructure is not apparent in our fiducial case. Our two test
  cases appear qualitatively similar, suggesting that the apparent
  substructure is an artificial effect induced by difference in the PSF.}\label{figure:appmaps}
\end{figure}

Maps of Pa$\alpha$ to H$\alpha$ for these three cases for galaxy G 20-2 are shown in Figure
\ref{figure:appmaps} alongside the HST map of H$\alpha$ without PSF
matching. The intrinsic value for this ratio assuming Case B
recombination is 0.123. Regions with Pa$\alpha$/H$\alpha$ greater than
indicate areas of nonzero $A_{H\alpha}$. The key difference between
the fiducial case and the two test cases is the presence of clumpy
substructure with the test cases showing a correspondence between
clump loci and minima in $A_{H\alpha}$. This is due to the lower
Strehl ratio of the OSIRIS PSF, with a large fraction of the light of
each spaxel being blurred to larger radii, resulting in an
artificially low Pa$\alpha$ to H$\alpha$ ratio. The fact that both
test cases exhibit this behavior gives us confidence that this is a
direct result of the difference in PSF shape between OSIRIS and HST.

\begin{figure}
  \includegraphics[width=\columnwidth]{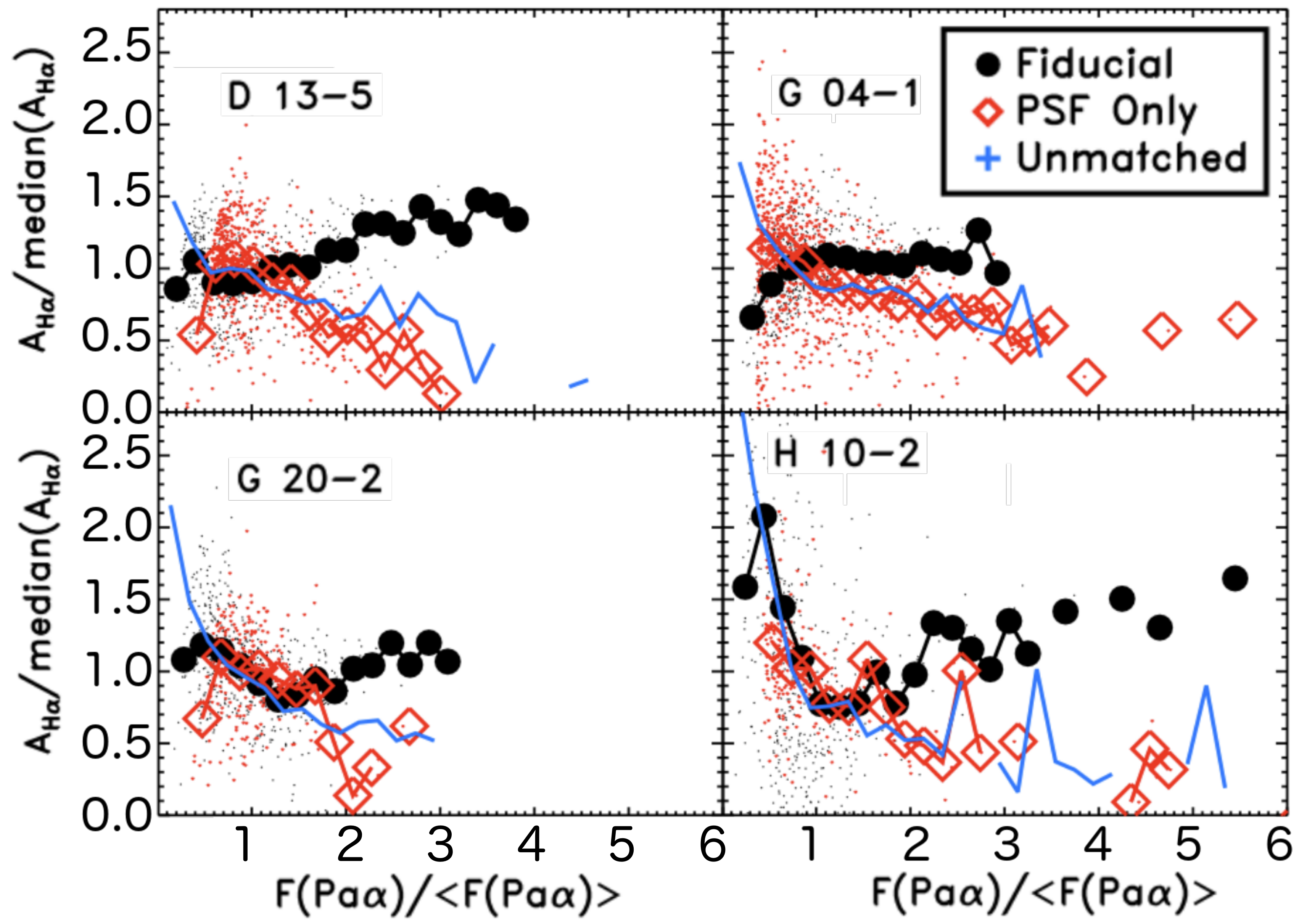}
  \caption{For each galaxy we compare $A_{H\alpha}$/median($A_{H\alpha}$) as a
    function of the Pa$\alpha$ flux in a given pixel normalized by the
  average flux for that galaxy. The small black points give individual
spaxel measurements for our observations and the large black circles
are binned values. Similarly we create mock observations described in
Section \ref{section:ch4psfmtest} using our PSF matched and non-matched
HST images, these values are plotted as red points for spaxels and red
diamonds for binned (bin sizes of 0.2 in $F($Pa$\alpha)$/$<F($Pa$\alpha)>$) values. Finally we plot as a solid blue line the
results using the observed Pa$\alpha$ and H$\alpha$ maps without PSF
matching. This non-PSF matched case is well matched to the PSF only
test case, thus our PSF matching procedure appears to accurately
capture the true difference in our observed PSFs.}\label{figure:ch4psfmtest}
\end{figure}

We plot the comparison between the three $A_{H\alpha}$ maps described above
in Figure \ref{figure:ch4psfmtest}. In each panel we plot the
normalised quantity, 
$A_{H\alpha}$/median($A_{H\alpha}$) as a function of
$F($Pa$\alpha)$/mean($F($Pa$\alpha)$). Values for the fiducial $A_{H\alpha}$ are given
by black points with large black circles indicating binned values (with bin
sizes of 0.2 in $F($Pa$\alpha)$/$<F($Pa$\alpha)>$). Red points
give the PSF Only values with open red diamonds indicating the binned
values. Finally, the thick blue line in each panel indicates the binned values
for the unmatched case. 

We find that for both the PSF Only and Unmatched cases $A_{H\alpha}$
decreases steadily with increasing Pa$\alpha$ flux. This is in
contrast to the roughly flat, or rising in the case of D 13-5 and H
10-2, behavior of our Fiducial case. This difference is entirely do to
the difference between the HST and OSIRIS PSF. The latter includes a
broad component, which is a common feature of AO observations, thus
distributing light from bright pixels into lower flux regions. In the
two cases with poorly matched PSFs (PSF Only and Unmatched) this
induces an artificial dearth of Pa$\alpha$ flux in high flux regions,
and, therefore, and under prediction of $A_{H\alpha}$. The fact that
our true unmatched images are well matched in behavior by the
simulated case using only the HST images (PSF Only) give us added
confidence that our OSIRIS PSF is a reasonable representation of the
response of our AO system to a point source. We note that Figure
\ref{figure:ch4psfmtest} shows a much larger spread for spaxels with
low Pa$\alpha$ flux and in the lowest flux bins all three test cases
seem to disagree. While this could be due to more intrinsic variation
in line-of-sight attenuation at low Pa$\alpha$ flux, more likely it
reflects spaxels that are near the detection limit as well as the
contamination of low flux spaxels by clump light blurred by the wings
of the AO-PSF. For this reason we stress some caution in
over interpretation of $A_{H\alpha}$ in low flux regions. The large
spread for individual spaxels also suggests that individual spaxel
measurements of $A_{H\alpha}$ are not entirely trustworthy, thus the
value of maps presented in Figure \ref{figure:dustmaps} is in the
average behavior of $A_{H\alpha}$ in different spatial regions of the
observed galaxies. This will trace large scale (i.e. clump-to-clump)
variations in attenuation that can be indicative of regions of very
high dust column density such as dust lanes.

\begin{figure}
  \includegraphics[width=\columnwidth]{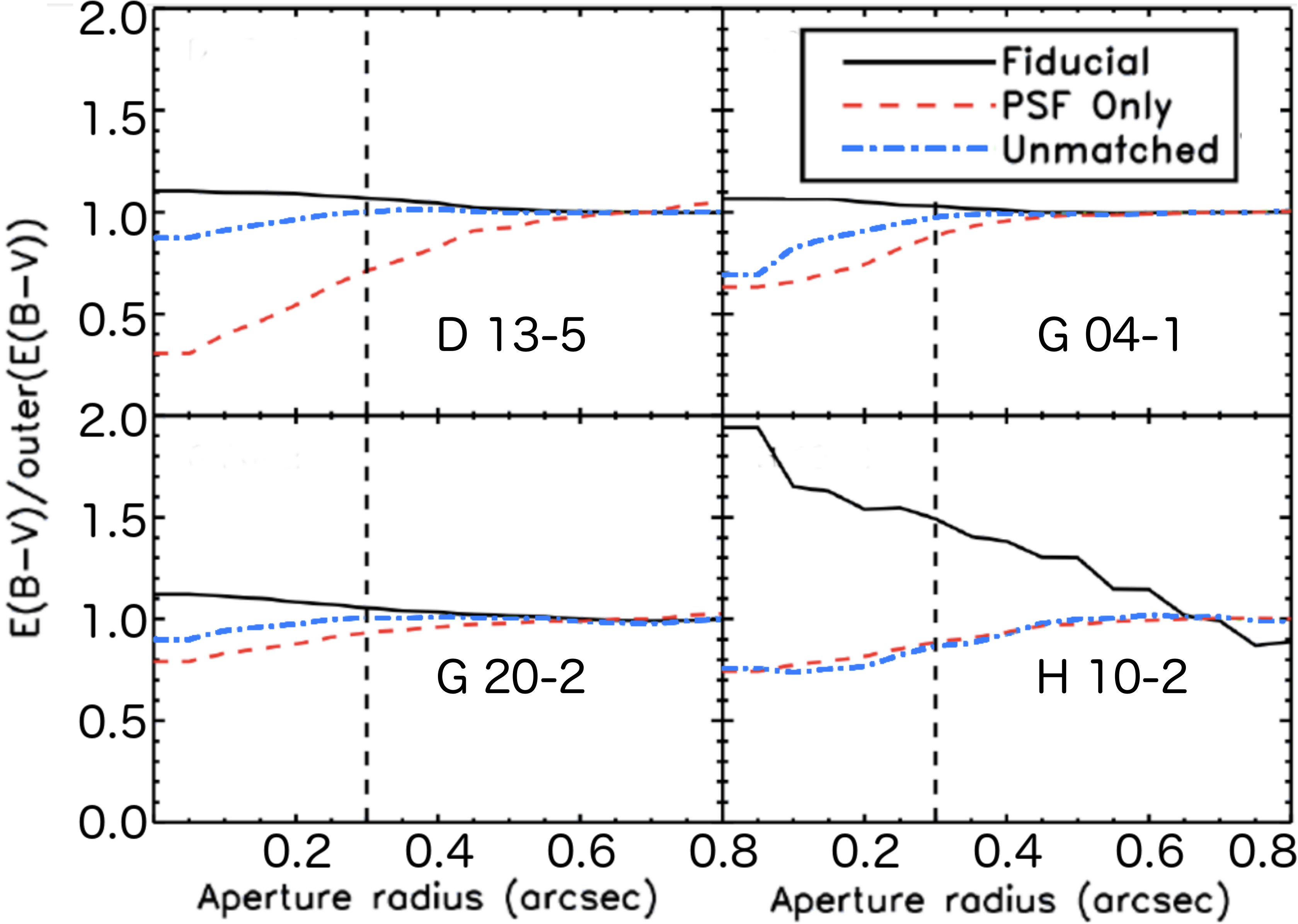}
  \caption{For each galaxy we measure $E(B-V)_{gas}$ in apertures of
    increasing radii for our PSF matched data, as well as the two test
  cases described in Section \ref{section:ch4psfmtest}, both of which do not
  include PSF matching. Here we plot $E(B-V)_{gas}$ as a function of
  aperture size averaged over all of the clumps within a given
  galaxy. Individual profiles are found to be less regular due
  variation induced by noise. We normalize each profile by the value in the largest
  apertures prior to averaging to account for absolute differences
  between each clump. For all galaxies we find the test cases exhibit
  a rising profile, a signature of the PSF mismatch, while our
  observations show a decrease. In each panel we plot a vertical
  dashed line at 0$\farcs$3, which roughly marks the transition at
  which the test case using both HST and OSIRIS data comes intro
  agreement with our observations. We note that this is typically
  quite close ($\sim$5\%) from the value in large apertures,
  reflecting the limitations imposed by the broad wings of the AO PSF.}\label{figure:ch4psfmap}
\end{figure}

We next test how this will affect integrated (aperture) measurements of $E(B-V)$ for
individual clumps in each galaxy. For clump measurements we sum
pixels/spaxels in apertures centered on each clump and compute the ratio of Pa$\alpha$ to H$\alpha$ in
these apertures. For each clump  we measure this ratio
in apertures of increasing size for our fiducial measurement
as well as the two test cases described above. We then compute
a value of $E(B-V)$ assuming a \citet{card89} attenuation curve. We normalize the $E(B-V)$ profiles
produced in this way by the value measured in large apertures
in order to compare results from clump to clump. The results of this
test are plotted in Figure \ref{figure:ch4psfmap}
with a panel for each galaxy. Profiles presented in this Figure
represent the average clump profile in a given galaxy. The
black solid line gives our fiducial measurements while red and green
lines give our PSF Only and Unmatched match test cases
respectively. In each panel the profiles are normalised to the value
measured for very large apertures that effectively encompass the
entire galaxy. Values in these large apertures will be negligibly
altered by any PSF effects as in all three cases all of the observed
emission from clumpy and non-clumpy regions will be included.

For both of our test cases we find a rising profile of $E(B-V)$ with
aperture size. This is expected considering the AO PSF, which will redistribute
Pa$\alpha$ flux from the centre of a given clump to
larger radii, and in agreement with maps of Pa$\alpha$/H$\alpha$
shown in Figure \ref{figure:appmaps}. HST photometry of H$\alpha$ does not suffer from this
effect and clumps observed at H$\alpha$ will be more centrally
concentrated. For our fiducial case, for which PSF matching has been
performed,
we find that $E(B-V)$ decreases with aperture radius in
all cases. We conclude that, because our unmatched and PSF only test
cases seem to typically exhibit similar behavior, our PSF matching
procedure provides a reasonable match between our two datasets. We also find that $r=0\farcs3$ apertures are
satisfactory and adopt these for measuring $A_{V}$ in DYNAMO clumps in
Section \ref{section:ch4clmeas}. This is not unreasonable as apertures
of this size are comparable to the core size of the OSIRIS AO PSF,
thus encompassing a majority of the well resolved portion of the clump
light. 

\section{GMOS Image Registration}\label{section:ch4imreg}

\begin{figure*}
  \centering
  \includegraphics[width=\textwidth]{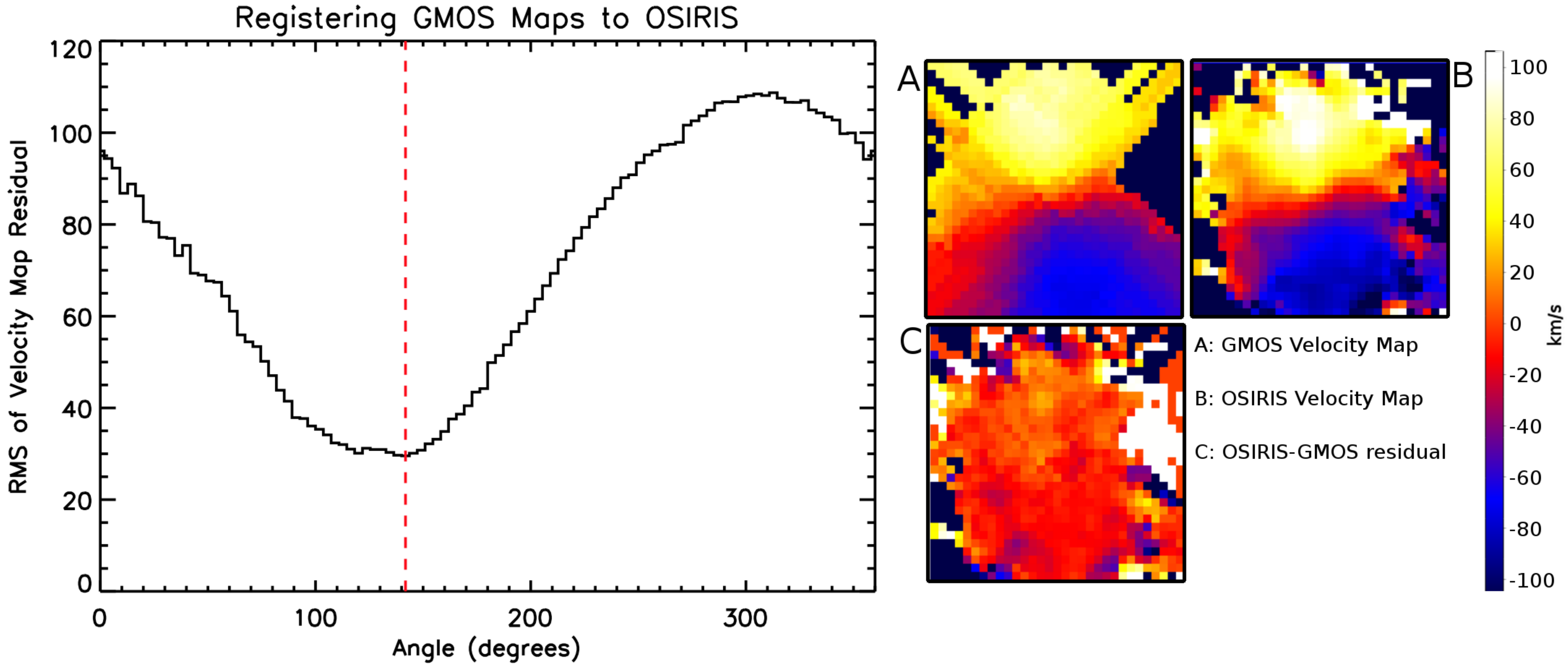}
  \caption{\textbf{left:} diagnostic plot of our GMOS registration
    procedure. Briefly we clip each image and match the galaxy centre
    based on the continuum image. The standard deviation of the values
  in the velocity map residual (OSIRIS$_{vm}$-GMOS$_{vm}$) is measured
for all angles and the necessary rotation is taken as the minimum of
the distribution, marked with a red dashed line. \textbf{right:}
examples of clipped images used to produce the left plot. Panel A
shows the GMOS velocity map rotated by the angle indicated by the red
dashed line, panel B shows the OSIRIS velocity map, and panel C shows
the residual velocity map. All three maps are shown using the same
colour scaling.}\label{figure:ch4gmosreg}
\end{figure*}

We register maps produced from our GMOS IFS
observations \citep[see][]{bassett14} to those from OSIRIS based on
the kinematic maps produced from the two datasets. Due to the lower
spatial resolution of the GMOS observations
the clumps begin to blend together hampering our ability to use the
registration method described for HST. In this case we wish to match
maps from two IFS datasets, both of which can provide a wealth of
spatially resolved information. We perform our registration in two
steps, the first of which is to match the centroid position of the
continuum image of the galaxy for both instruments. We then shift the
velocity and velocity dispersion maps produced from our GMOS
observations based on this shift and produce rotated versions of these
kinematic at a large number of rotation angles locating the angle
that minimizes the difference average residual in both velocity
($v_{OSIRIS}-v_{GMOS,rotated}$) and velocity dispersion
($\sigma_{OSIRIS}-\sigma_{GMOS,rotated}$). We also perform the test
adjusting the x and y shifts estimated based on the continuum images
by $\pm$0$\farcs$3 to account for the seeing of our GMOS
observations. An example of this
procedure is shown in Figure \ref{figure:ch4gmosreg}. 

After selecting the
best fitting shift and rotation angle, these are applied to the maps
of  [OIII]/H$\beta$, giving a more complete understanding of the
state of the gas in DYNAMO HST+OSIRIS galaxies. In particular,
variations in [OIII]/H$\beta$ may indicate the presence of
ionizing processes other than star-formation that may be responsible
for a portion of the ionizing radiation. In this case, Hydrogen line
ratios which we use to estimate the amount of attenuation suffered by
ionized gas, may no longer follow the relation given by Case B
recombination resulting in a poor estimate of $A_{V}$. Registered maps of [OIII]/H$\beta$ from GMOS
to our HST+OSIRIS maps of nebular attenuation in Figure
\ref{figure:ch4oratmaps} in Section
\ref{section:ch4resavres}. 


\bsp	
\label{lastpage}
\end{document}